\documentclass{article}
\usepackage[nonatbib,final]{neurips_2025}
\usepackage[utf8]{inputenc}
\usepackage[T1]{fontenc}
\usepackage{hyperref}
\usepackage{url}
\usepackage{booktabs}
\usepackage{amsfonts}
\usepackage{nicefrac}
\usepackage{microtype}
\usepackage{xcolor}
\usepackage{amsthm}
\usepackage{amsmath}
\usepackage[toc,page,titletoc]{appendix}
\usepackage{minitoc}
\usepackage{multirow}
\usepackage{graphicx}
\usepackage{subcaption}
\usepackage{xspace}
\usepackage{xcolor}
\usepackage{enumitem}
\usepackage{hyperref}
\usepackage{listings}
\usepackage{cleveref}
\usepackage{wrapfig}

\usepackage{colortbl}
\usepackage{amssymb}
\usepackage{algorithm}
\usepackage{algorithmic}

\usepackage{makecell}
\usepackage{tabularx}

\usepackage[framemethod=TikZ]{mdframed}

\usepackage{pgfplots}
\usepackage{pgfplotstable}
\pgfplotsset{compat=1.18}

\usepackage{booktabs}
\usepackage{caption}
\usepackage{adjustbox}

\usepackage{booktabs}
\usepackage{enumitem}
\usepackage{multirow}
\usepackage{xcolor}
\usepackage{amsthm}
\newtheorem{lemma}{Lemma}[section]
\usepackage{pifont}

\title{\sreagent{}: A Multi-agent System for Autonomous Reliability Engineering of Modern Clouds}

\author{%
  Yinfang Chen$^{\diamond}$\thanks{Co-primary authors.} \and 
  \textbf{Jiaqi Pan$^{\diamond\P*}$} \and 
  \textbf{Jackson Clark$^{\diamond*}$} \and  
  \textbf{Yiming Su$^{\diamond*}$} \and 
  \textbf{Noah Zheutlin$^\dag$} \and       
  \textbf{Bhavya Bhavya$^\dag$} \and    
  \textbf{Rohan Arora$^\dag$} \and      
  \textbf{Yu Deng$^\dag$} \and 
  \textbf{Saurabh Jha$^\dag$} \hspace{2cm}             
  \textbf{Tianyin Xu$^{\diamond}$} \\
  University of Illinois Urbana-Champaign$^{\diamond}$\ \ \ IBM Research$^\dag$\ \ \ IIIS, Tsinghua University$^{\P}$
}

\definecolor{bg}{rgb}{0.98, 0.98, 0.98} 

\AtBeginDocument{%
  \providecommand\BibTeX{{%
    \normalfont B\kern-0.5em{\scshape i\kern-0.25em b}\kern-0.8em\TeX}}}
    
\usepackage{cite}
\usepackage{tcolorbox}
\definecolor{bg}{gray}{0.95}
\definecolor{darkred}{RGB}{139, 0, 0} % Dark red
\tcbuselibrary{breakable,xparse,skins,listings}
\DeclareTCBListing{mintedbox}{O{}m!O{}}{%
  breakable=true,
  listing engine=listings,
  listing only,
  listing options={%
    language=#2,
    basicstyle=\small\ttfamily,
    numbers=left,
    numbersep=8pt,
    breaklines=true,
    gobble=0,
    #1},
  boxsep=0pt,
  left skip=2pt,
  right skip=0pt,
  left=25pt,
  right=0pt,
  top=3pt,
  bottom=3pt,
  arc=5pt,
  leftrule=0pt,
  rightrule=0pt,
  bottomrule=2pt,
  toprule=2pt,
  colframe=orange!80,
  enhanced,
  overlay={%
    \begin{tcbclipinterior}
    \fill[orange!20!white] (frame.south west) rectangle ([xshift=20pt]frame.north west);
    \end{tcbclipinterior}},
  #3}
\colorlet{shadecolor}{gray!30}
\definecolor{Gray}{gray}{0.8}

\newenvironment{packed_itemize}{
\begin{list}{\labelitemi}{\leftmargin=1.0em}
 \setlength{\itemsep}{2.5pt}
 \setlength{\parskip}{0pt}
 \setlength{\parsep}{0pt}
 \setlength{\headsep}{0pt}
 \setlength{\topskip}{0pt}
 \setlength{\topmargin}{0pt}
 \setlength{\topsep}{0pt}
 \setlength{\partopsep}{0pt}
}{\end{list}}

\newtheoremstyle{findingstyle}
  {3pt}   % ABOVESPACE
  {3pt}   % BELOWSPACE
  {}  % BODYFONT
  {0pt}       % INDENT (empty value is the same as 0pt)
  {\bfseries} % HEADFONT
  {.}         % HEADPUNCT
  {5pt plus 1pt minus 1pt} % HEADSPACE
  {}          % CUSTOM-HEAD-SPEC

\theoremstyle{findingstyle}
\newtheorem{findinner}{\textbf{Finding}}

\newenvironment{find}
  {\begin{shaded}\begin{findinner}}
  {\end{findinner}\end{shaded}}

\newcommand{\sreagent}{\textsc{Stratus}}

\newcommand{\code}[1][]{\lstinline[language=Python,basicstyle=\footnotesize\ttfamily,#1]}

\def\Snospace~{\S{}}

\definecolor{mellowgreen}{RGB}{200,230,200}

\newtcblisting{agenttask}[2]{
    listing only,
    listing engine=listings,
    colframe=cyan!40!black,colbacktitle=cyan!20,
    coltitle=black,boxrule=1.2pt,enhanced,arc=2pt,fonttitle=\bfseries\scriptsize,
    title={\small Task: {#1}\hfill {#2} Agent},
    listing options={basicstyle=\tiny\ttfamily,breaklines=true,gobble=0},
    breakable
}

\newtcblisting{agentanswer}[2]{
    listing only,
    listing engine=listings,
    colframe=cyan!40!black,colbacktitle=cyan!20,
    fontupper=\ttfamily\tiny,
    coltitle=black,boxrule=1.2pt,enhanced,arc=2pt,fonttitle=\bfseries\scriptsize,
    title={\small Task {#1} Finished \hfill {#2} Agent},
    listing options={basicstyle=\tiny\ttfamily,breaklines=true,gobble=0},
    breakable
}

\newtcolorbox{agentthought}[1]{
    colframe=lime!40!black,colbacktitle=lime!20,
    fontupper=\ttfamily\tiny,
    coltitle=black,boxrule=1.2pt,enhanced,arc=2pt,fonttitle=\bfseries\scriptsize,
    title={\small Thought \& Action \hfill {#1} Agent},
    breakable
}

\newtcolorbox{agentgoldthought}[2]{
    colframe=yellow!70!black,colbacktitle=yellow!60,
    fontupper=\ttfamily\tiny,
    coltitle=black,boxrule=1.2pt,enhanced,arc=2pt,fonttitle=\bfseries\scriptsize,
    title={\small Gold Action {#1} \hfill {#2} Agent},
    breakable
}

\newtcblisting{agenttool}[2]{
    listing only,
    listing engine=listings,
    colframe=yellow!40!black,colbacktitle=yellow!20,
    fontupper=\ttfamily\tiny,
    coltitle=black,boxrule=1.2pt,enhanced,arc=2pt,fonttitle=\bfseries\scriptsize,
    title={\small Tool Usage: {#1}\hfill {#2} Agent},
    listing options={basicstyle=\tiny\ttfamily,breaklines=true,gobble=0},
    breakable
}

\newtcblisting{frameworklog}[2]{
    listing only,
    listing engine=listings,
    colframe=magenta!40!black,colbacktitle=magenta!20,
    fontupper=\ttfamily\tiny,
    coltitle=black,boxrule=1.2pt,enhanced,arc=2pt,fonttitle=\bfseries\scriptsize,
    title={\small Phase: {#1}\hfill Run \#{#2}},
    listing options={basicstyle=\tiny\ttfamily,breaklines=true,gobble=0},
    breakable
}

\begin{document}

\maketitle

\begin{abstract}
In cloud-scale systems, failures are the norm.
A distributed computing cluster exhibits hundreds of machine failures 
    and thousands of disk failures; software bugs and misconfigurations
    are reported to be more frequent. 
The demand for {\it autonomous}, AI-driven reliability engineering continues to grow,
    as existing human-in-the-loop practices
    can hardly keep up with the scale of modern clouds.
This paper presents \sreagent{}, 
    an LLM-based multi-agent system for realizing autonomous Site Reliability Engineering (SRE) of cloud services.
\sreagent{} consists of multiple specialized 
    agents (e.g., for failure detection, diagnosis, mitigation), organized in
    a state machine to assist system-level safety reasoning 
    and enforcement.
We formalize a key safety specification of agentic SRE systems like \sreagent{}, 
    termed {\it Transactional No-Regression (TNR)}, which 
    enables safe exploration and iteration.
We show that TNR can effectively improve
    autonomous failure mitigation. 
\sreagent{} significantly outperforms state-of-the-art SRE agents
    in terms of success rate of failure mitigation problems in AIOpsLab and ITBench 
    (two SRE benchmark suites), by at least 1.5 times across various models.
\sreagent{} shows
    a promising path toward practical deployment of agentic systems for cloud reliability. 
\end{abstract}

\section{Introduction}

Cloud systems are the backbone of today's large-scale computing and end-user applications.
Their reliability is crucial, with a few minutes of outages leading to significant user impacts
    and financial losses~\cite{Barroso:2018,Armbrust:09}.
Given the ever-growing scale, complexity, and dynamics of cloud systems,
    cloud reliability continues to be a grand challenge.
Despite extensive efforts~\cite{leners:11,Huang:2018,ren:osdi:23,Zhang:17},
    modern clouds still struggle to detect, localize, and mitigate system failures effectively and timely.
One essential reason is the existing human-centric reliability engineering practice~\cite{Beyer:2016,Beyer:2018}---human engineers are 
\begin{figure}[htbp]
    \centering
    \vspace{-5pt}
    \includegraphics[width=0.975\textwidth]{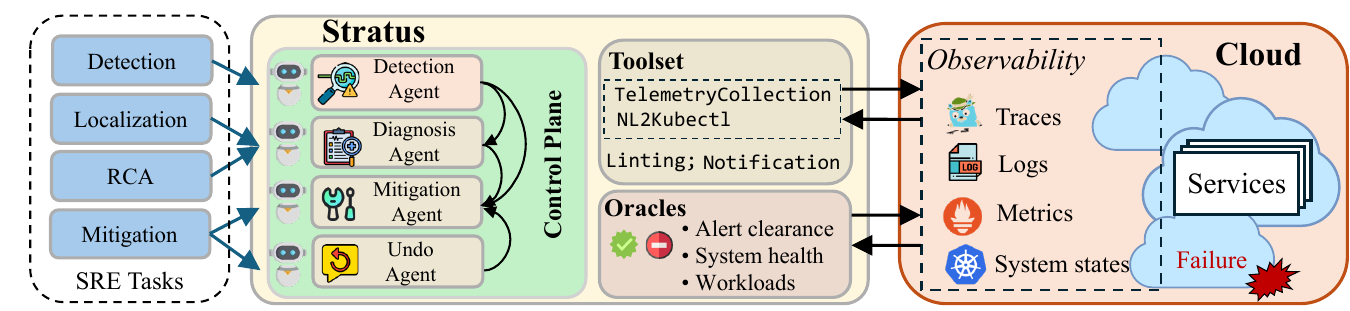}
    \caption{Overview of \sreagent{}, an LLM-based multi-agent system for autonomous Site Reliability Engineering (SRE) of modern cloud services.}
    \label{fig:agent-arch}
\end{figure}
in the loop,
    responsible for decision making, e.g., which alerts to act upon,
    which tools to use, and which mitigation paths to choose.
However, as failures are the norm in cloud-scale systems,
    it is increasingly costly for human-in-the-loop approaches 
    to keep up with the scale of modern clouds.
Autonomous Site Reliability Engineering (SRE), empowered by generative AI, is desired~\cite{shetty2024building,jha2025itbench,chen2025aiopslab}.

This paper explores an agentic AI approach for {\it autonomous} Site Reliability Engineering (SRE) of modern clouds.
We present \sreagent{}, 
    an LLM-based agentic system that realizes autonomous
    SRE for cloud services through
    failure detection, localization, root-cause analysis, and mitigation.
\sreagent{} differs from prior work on developing AI or agentic tools that 
    assist human engineers~\cite{chen:eurosys:24,jiang2024xpert,jin2023assess,shetty2022autotsg,wang2023rcagent,rcaflash}
    (for data collection, summarization, and root-cause prediction).
\sreagent{} aims to govern live production systems and mitigate failures, without 
    human intervention (see \S\ref{sec:background}).

\sreagent{} is a multi-agent system that orchestrates specialized 
    agents of detection, diagnosis, mitigation, etc., 
    each responsible for one or multiple tasks (Figure~\ref{fig:agent-arch}).
We chose a multi-agent design because 
    (1) it specializes agents for SRE tasks with well-defined roles,
    (2) it offers customizability of individual agents 
        and extensibility of the system,
    and (3) it makes it easy to reason about agent safety.
\sreagent{} uses deterministic control-flow logic that orchestrates
    agents in a state machine, while leveraging LLMs in data flows that make agents intelligent and creative.

Safety is a key challenge of SRE agents (or any agentic systems that operate critical live systems).
\sreagent{} must not worsen the state of the target system, 
    when mitigating its failures; however, it is hard to ensure that a mitigation plan 
        generated by the agent (or even by a human)
    is always successful, considering the dynamics and complexity of the system 
        and imperfect AI models. 
In practice, the risk of ``making things
    worse'' is a fundamental barrier to deploying autonomous agentic techniques in high-stake systems.
We formalize a safety specification of agentic SRE, termed {\it Transactional No-Regression} or TNR.
TNR constructs transaction semantics to ensure:
    (1) the agent's mitigation actions, if unsuccessful, can always be ``undone,''
    and (2) the agent keeps improving the system in terms of its health state,
    by undoing actions that worsen it (i.e., no regression).
We show that TNR enables safe agent exploration and iteration, which effectively improves failure mitigation.

We describe our implementation that realizes TNR using (1) state-machine based scheduling
    (for write exclusivity),
    (2) sandboxing (for confinement), 
    and (3) an {\it Undo Agent} that realizes a system-wide undo operator~\cite{undo}.
We implement a stack-based undo mechanism that 
    tracks the agent actions with regards to specific system state
    and reverts them in the correct order if needed.
We also describe other notable implementations, e.g., agent tooling,
    bootstrapping, and termination.

We evaluate \sreagent{} on two SRE benchmark suites, 
    AIOpsLab~\cite{chen2025aiopslab} and ITBench~\cite{jha2025itbench}.
\sreagent{} significantly outperforms state-of-the-art agentic solutions 
    in terms of success rate of failure mitigation
    problems in AIOpsLab and ITBench, by at least 1.5 times across various models (GPT-4o, GPT-4o-mini, and Llama3).
We show that TNR, which enables the undo-and-retry practice, is
    key to the ability of \sreagent{} in solving complex mitigation problems; it allows
    \sreagent{} to safely and iteratively explore new mitigation paths, 
    while avoiding unrecoverable error states.

{\bf Summary.} This paper makes the following contributions: 
\begin{packed_itemize}
    \item An attempt of an agentic AI system that autonomously mitigates system failures,
        towards fully autonomous Site Reliability Engineering of modern clouds.
    \item \sreagent{}, a multi-agent system that implements an end-to-end agentic SRE pipeline, from failure detection,
        localization, to mitigation, with system-wide safety properties. 
    \item Transactional Non-Regression (TNR), a formalization of agentic SRE safety specification, 
        and its implementation in \sreagent{} that
        enables safe agent exploration and iteration.
    \item \sreagent{} significantly outperforms state-of-the-art SRE agents in terms of success rate of 
        various SRE tasks on two SRE benchmark suites (AIOpsLab and ITBench). 
    \item The artifacts of \sreagent{} can be found at 
    {\small \url{https://github.com/xlab-uiuc/stratus}}.
    % {\small \url{https://anonymous.4open.science/r/stratus-agent}}.
\end{packed_itemize}

\section{Background and Problem Definition}
\label{sec:background}

\sreagent{} aims to realize {\it autonomous SRE} which we define 
    as an automated system on detecting failures, 
    localizing failing components, analyzing the root causes, and mitigating the failures,
    with minimal human intervention.
In other words, autonomous SRE systems like \sreagent{} imitate human engineers with agentic AI,
    and hence is fundamentally different from AI or agentic tools designed
    for assisting human engineers on specific SRE tasks~\cite{wang2023rcagent,chen:eurosys:24,rcaflash}.
A key differential feature of \sreagent{} is to automatically {\it mitigate} failures, beyond summarizing 
    failure information and predicting root cause categories as in prior work.
Mitigation poses unique challenges for autonomous SRE: unlike detection and diagnosis tasks
    that only need to observe the system,
    mitigation needs to change the system state.
How to ensure safety of failure mitigation is a key challenge---mitigation actions must 
    not cause new (nested) failures and must not drive the system to a worse state (i.e., no regression). 

We model a target cloud system as an {\it environment} $\mathcal{E}$ characterized by:
\begin{packed_itemize}
    \item A set of states $S = \{s\}$, including a crash state $\bot$ where the system is completely unavailable;
    \item A severity metric $\mu (s^e): S \cup \{\bot\} \to \mathbb{N} \cup \{\infty\}$ that represents the severity of an error state $s^e$.
\end{packed_itemize}
An error state can be caused by different types of faults (root causes)~\cite{aviz:04},
    including software bugs~\cite{Gunawi:14}, misconfigurations~\cite{xu:13}, and 
    hardware issues~\cite{Hochschild:2021, meza:imc:18, maneas2020ssdreliability,meza2015flashmemoryfailure}.
The severity of the error for a given state $s$ 
    is measured as a weighted sum of 
    the sets of alerts ($A$), violations of service-level agreement (SLA) ($V$), 
    and system capacity loss (unhealthy nodes) ($L$);
    it is formally defined as:
    $\mu(s)=w_1 \cdot |A| + w_2 \cdot |V| +
    w_3 \cdot |L|$ where $w_i>0, \mu(\bot)=\infty$. 
If the entire system is unavailable, then it is in the state $\bot$, where $\mu(\bot) = \infty$.
Just like human SRE engineers~\cite{Beyer:2016}, 
\sreagent{} addresses following SRE tasks 
    in order to resolve the error in the system state:
\begin{packed_itemize}
    \item {\bf Detection.} SRE must promptly detect production failures via logs, traces, and other telemetry data; 
        detecting failures is the first step to prevent incidents.
    \item {\bf Localization.} SRE must localize the faulty components so that they can be 
        isolated to minimize the blast radius and impacts.
    \item {\bf Root Cause Analysis (RCA).} SRE shall identify the root causes of the failures in order to 
        repair the faulty components (e.g., fixing bugs and misconfigurations).  
    \item {\bf Mitigation.} SRE must mitigate the failures to prevent their propagations that 
        lead to production incidents and service outages. 
\end{packed_itemize}
Note that RCA is not on the critical path of mitigation. In practice, RCA is typically done offline,
    while mitigation directly determines service availability~\cite{Treynor:17}.
A successful mitigation may not need to know the failure root causes.
For example, mitigating a faulty machine can be done by migrating deployed jobs 
    to a healthy machine~\cite{veeraraghavan:osdi:18}, without knowing the faulty component;
    mitigating a faulty software component can be done by rebooting it~\cite{qin:05,candea:04}
    or reverting recent changes~\cite{Bhagwan:osdi:18}.
\section{\sreagent{} Design}
\label{sec:sreagent}

\sreagent{} uses a multi-agent system design that orchestrates multiple specialized agents 
    $\mathcal{A} = \{ \alpha_i \}$ ($\alpha_i$ refers to an agent) 
    to interact with $\mathcal{E}$ and mitigate any system failure represented by an error state $s^e$. 
Our implementation currently uses four agents $\mathcal{A} = \{ \alpha_D, \alpha_G, \alpha_M, \alpha_U \}$:
\begin{packed_itemize}
    \item $\alpha_D$ ({\bf Detection agent}): 
        Observes $\mathcal{E}$ to identify failures and establishes the initial error state $s_0^e$.
    \item $\alpha_G$ ({\bf Diagnosis agent}):
         Observes $\mathcal{E}$ ($s_0^e$ and telemetry) to determine the root cause(s) of the detected failure; 
         its output is primarily analytical. $\alpha_G$ is responsible for both localization and RCA.
    \item $\alpha_M$ ({\bf Mitigation agent}): Takes the diagnostic output from $\alpha_D$ and $\alpha_G$. 
        It is responsible for (1) devising a high-level mitigation plan,
        (2) decomposing the plan into a sequence of concrete mitigation actions, 
        and (3) executing each action by issuing commands through agent tools with Agent-Computer Interfaces (ACI)~\cite{yang2024swe} (see \S\ref{sec:impl}). 
    \item $\alpha_U$ ({\bf Undo Agent}): Executes an undo sequence on $\mathcal{E}$ if a transaction executed by $\alpha_M$ is aborted.
\end{packed_itemize}
Agents interact with the environment using actions from a defined {\it Action Space} through ACI:
\begin{packed_itemize}
    \item $A_{\text{read}}$: Read-only commands that do not change the state of $\mathcal{E}$ (e.g., \texttt{kubectl} \texttt{get}\footnote{\texttt{kubectl} is 
        the command-line tool of Kubernetes, a {\it de facto} cloud platform that manages cloud systems~\cite{Burns:2016}.}).
    \item $A_{\text{write}}$: Write commands that can mutate the state of $\mathcal{E}$ (e.g., \texttt{kubectl} \texttt{apply}).
    \item $A_{\text{undo}}$: A specialized sequence of commands executed by $\alpha_U$ to undo the effects of write commands.
\end{packed_itemize}
Agents $\alpha_D, \alpha_G$ are restricted to $A_{\text{read}}$. 
Agent $\alpha_M$ executes actions from $A_{\text{write}}$, as well as $A_{\text{read}}$ for reading system states
    and checking pre/post conditions. 
Agent $\alpha_U$ executes actions effectively in $A_{\text{undo}}$ (which internally uses $A_{\text{write}}$ commands to restore a prior state).

The system state of $\mathcal{E}$ changes based on {\it State-Transition Relations} (see~\cite{gu:sosp:23}),
    $R : S \times A_{\text{write}} \to S \cup \{\bot\}$. 
The undo is performed by an {\it operator} $U : S \to S$, realized by $\alpha_U$.

\subsection{Safety Specification}
\label{sec:safety}

\sreagent{} must guarantee safety.
However, an agent's action, whether due to an imperfect understanding, 
    a planning flaw, or hallucinations of the generative model, 
    could inadvertently worsen an already degraded system state ($\mu(s_{0}^e) > 0$). 
In our experience, this risk of ``making things worse'' is a main 
    barrier to deploying autonomous agentic techniques in high-stake systems.
How can we grant autonomous agents the authority to execute sequences of potentially 
    state-changing commands (e.g., via \texttt{kubectl})
    for critical tasks like failure mitigation, without risking worse outcomes?
    
We formalize a safety specialization of agentic SRE systems like \sreagent{}, 
    termed Transactional Non-Regression (TNR), 
    atop the classic transition abstraction~\cite{haerder1983transactionorientedrecovery}.
TNR ensures that
    (1) the agent's mitigation actions, if unsuccessful, can always be ``undone'',
    and (2) the agent keeps improving the system state (i.e., the severity metric 
    decreases monotonically, with no regression).
\subsubsection{Assumptions}
\label{subsec:assumptions}
We state the following assumptions, which are enforced by our implementation (\S\ref{sec:impl}):
\begin{enumerate}[label=A\arabic*,leftmargin=*,itemsep=0pt]
\item \textbf{Writer Exclusivity (Agent-Lock or A-Lock).} At most one writer agent ($\alpha_M$ or $\alpha_U$) 
    is scheduled to execute commands that can mutate system states at a time and have exclusive access;
    reader agents ($\alpha_D, \alpha_G$) do not mutate state relevant to $\mu$.
    A-Lock is a readers-writer lock.
\item \textbf{Faithful Undo.} The undo operator $U$, when invoked by $\alpha_U$ on $s_{\text{post}}$ after an abort decision, restores the checkpointed $s_{\text{pre}}$ exactly, i.e., $U(s_{\text{post}})=s_{\text{pre}}$.
\item \textbf{Bounded Risk Window.} The length $k$ (the number of commands) of any transaction executed by $\alpha_M$ 
    is bounded by a system-wide threshold $K$ ($1 \le k \le K$). This limits the duration the A-Lock is held and the complexity of any single transaction.
\end{enumerate}
\subsubsection{Transaction Semantics} 
\label{subsec:mt}

We construct transaction semantics atop the above assumptions, where a transaction 
    of length $k$ $(1\!\le\!k\!\le\!K)$ is
    a sequence of read or write commands $T=(a_{1},\dots,a_{k}) \in (A_{\text{write}}, A_{\text{read}})^k$. 
%In \sreagent{}, 
A transaction encodes a mitigation plan executed 
    by $\alpha_M$ while holding the A-Lock (see A1):
\begin{enumerate}[leftmargin=*,label=\textit{R\arabic*}] 
\item \textbf{Checkpoint}\,: $\alpha_M$ records the entry state $s_{\text{pre}}$ before $a_1$ is executed.
\item \textbf{Execute}\,: $\alpha_M$ runs $a_{1}\ldots a_{k}$ sequentially. Let $s_{\text{post}}$ be the state after $a_k$ completes (or $\bot$ if any $a_j$ $(1\!\le\!j\!\le\!k)$ causes a crash).
\item \textbf{Commit/Abort rule}\,:\;
      $\alpha_M$ evaluates: \emph{Commit} if $s_{\text{post}}\neq\bot$ and
      $\mu(s_{\text{post}})\le\mu(s_{\text{pre}})$;
      otherwise, \emph{abort} by instructing $\alpha_U$ to invoke $U$ \emph{once}. Per A2, this restores $s_{\text{pre}}$ exactly.
\end{enumerate}
An aborted transaction leaves \emph{no trace} on externally visible state transitions. 
We call the internal sequence of states visited during transaction execution, 
    $s_{\text{pre}} \xrightarrow{a_1} s^{(1)} \xrightarrow{a_2} \dots \xrightarrow{a_k} s_{\text{post}}$, 
    and their corresponding severity metrics 
    the \textit{hidden $\mu$-path}. Only $\mu(s_{\text{pre}})$ and, if $T$ commits, 
    $\mu(s_{\text{post}})$ are part of the externally \textit{visible} state transitions 
    that define the system's trajectory viewed by other agents.
\subsubsection{\bf Transactional Non-Regression (TNR)}
\label{subsec:tnr}
Let $s_{0}^e$ be the error state observed by $\alpha_D$ when an failure occurs,
    with an
\emph{initial severity} $b=\mu(s_{0}^e)$.

\begin{lemma}
Every state $s$ in externally visible state transitions (i.e., $s=s_0^e$, 
    or $s$ is a state immediately following a read-only action by $\alpha_D$ or $\alpha_G$, or $s$ is a state following the completion 
    (commit or abort) of a transaction by $\alpha_M or \alpha_U$) satisfies $\mu(s)\le b$.
\end{lemma}
\begin{proof}[\textbf{Proof Sketch.}]
The proof is by induction on the sequence of externally visible states, 
    where an externally visible state is one observed outside of any active transaction $T$.

\textit{Base Case.} For the initial state $s_{0}^e$ (established by $\alpha_D$ when the failure is detected), $\mu(s_{0}^e) = b$ by definition. Thus, the invariant $I(s) \equiv \mu(s) \le b$ holds for $s_0^e$.

\textit{Inductive Step.} Assume that the invariant $I(s_i) \equiv \mu(s_i) \le b$ holds for an externally visible state $s_i$. We show it holds for the next externally visible state $s_{i+1}$.
\begin{packed_itemize}
    \item If $s_{i+1}$ results from a \emph{read} action (by $\alpha_D$ or $\alpha_G$) on $s_i$: These agents execute actions from $A_{\text{read}}$ which do not mutate state. Thus, $s_{i+1}=s_i$, and $\mu(s_{i+1}) = \mu(s_i) \le b$. The invariant holds.
    \item If $s_{i+1}$ results from the completion of $T$ executed by $\alpha_M$, starting from $s_i$ (so $s_{\text{pre}}=s_i$):
        \begin{packed_itemize}
            \item The $T$'s execution ($a_1, \dots, a_k$) and its commit/abort decision occur atomically with respect to other writer agents due to {\it A-Lock (A1)}.
            \item If the $T$ \emph{commits}, the new externally visible state is $s_{i+1} = s_{\text{post}}$. The commit condition requires $\mu(s_{\text{post}})\le\mu(s_{\text{pre}})$. Since $s_{\text{pre}}=s_i$ and $\mu(s_i) \le b$ by the inductive hypothesis, it follows that $\mu(s_{\text{post}}) \le b$. The invariant holds for $s_{i+1}$.
            \item If the $T$ \emph{aborts}, $\alpha_U$ restores $s_{\text{pre}}$ with {\it Faithful Undo (A2)}. The externally visible state is $s_{i+1} = s_{\text{pre}}$ (which is $s_i$). Since $\mu(s_i)\le b$ by the inductive hypothesis, 
                the invariant holds for $s_{i+1}$. Note that the effects of $a_1, \dots, a_k$ by $\alpha_M$ are not part of the externally visible sequence.
        \end{packed_itemize}
\end{packed_itemize}
\vspace{-10pt}
Thus, any state in the externally visible sequence satisfies $\mu(s)\le b$. 
\end{proof}

Table~\ref{tab:scenarios} gives an example that contrasts the \emph{hidden} and \emph{visible}
severity trajectories for four common mitigation plans, as orchestrated by \sreagent{}.

\begin{table}[h]
\centering
\footnotesize
\vspace{-15pt}
\caption{The TNR safety guarantee under four failure mitigation plans.
\label{tab:scenarios}}
% \vspace{0.75em}
\begin{tabular}{p{2.8cm} p{4.0cm} p{2.0cm} c  p{1.5cm}}
\toprule
Mitigation & TNR actions (by $\alpha_M$) &
Hidden $\mu$-path\footnotemark &
Commit? & Visible $\mu$ \\ \midrule
Node drain/rebalance &
{cordon}, {evict}, {scale} &
$12\!\rightarrow\!18\!\rightarrow\!9$ &
\textcolor{green}{\ding{51}} & 
$12\!\rightarrow\!9$ \\[2pt]
Rolling upgrade &
{scale 0}, {patch}, {scale 3} &
$15\!\rightarrow\!22\!\rightarrow\!11$ &
\textcolor{green}{\ding{51}} & 
$15\!\rightarrow\!11$ \\[2pt]
Bad image attempted & 
{scale 0}, {patch(bad)}, {scale 3} & 
$15\!\rightarrow\!24\!\rightarrow\!30$ &
\textcolor{red}{\ding{55}} & 
$15\!\rightarrow\!15$ \\[2pt]
Single hot-fix ($K{=}1$) &
{apply hotfix} &
$15\!\rightarrow\!x$ &
\makecell{\textcolor{green}{\ding{51}} if $x\!\le\!15$\\\textcolor{red}{\ding{55}} otherwise} & 
$\le 15$ \\
\bottomrule
\end{tabular}
\vspace{-0.5cm}
\end{table}
\footnotetext{Hidden $\mu$ values arise \emph{inside} the transaction during execution by $\alpha_M$ and are not
visible to other agents, nor do they form part of the externally visible state sequence against which the lemma is proven.}

\subsection{Implications}
\label{sec:model:discussion}

TNR, which states that ``severity never increases over the initial baseline $b$ in the observable trace of states,'' 
    is an instance of the \emph{Alpern-Schneider safety property} \cite{alpern1987safety,alpern1986safetywithoutstuttering}.
This safety property is not merely a theoretical construct; 
    it forms a foundation for building trustworthy and effective AI agents
    that interact with critical live systems. 
For any agentic system employing an ACI to enact changes, 
    especially those using LLMs for complex reasoning and planning (like \sreagent{}'s $\alpha_M$), 
    the risk of generating sub-optimal or flawed action sequences is non-trivial. 

By ensuring that the potential ``damage''
    of an agent is capped (it cannot make the system observably worse than $s_0$), 
    TNR provides a powerful tool for \emph{safe exploration and iteration}. 
This allows the agent to attempt ambitious, complex repairs, learn from outcomes (via observing $\mu$ post-transaction), 
    and adapt its strategy without the risk of digging a deeper hole.

Moreover, preventing agent-driven escalations beyond the baseline $b$ 
    makes the \emph{liveness} property of eventually reaching a healthy state tenable~\cite{alpern1985liveness,alpern1987safety}. 
It empowers robust, reliable AI agents capable of sophisticated, autonomous interventions in the real world, 
    fostering the development of AI systems that can not only act, but also act \emph{safely} through {\it recoverability}.

\subsection{Extension}    

TNR currently assumes no concurrent writer agents and has no concurrency control.
    Writer agents are strictly serialized with the readers-writer lock (A1).
Our current deployment model uses one \sreagent{} instance per cloud system
    (each system managed in a separate namespace).
% We discuss extensions to support multiple concurrent writers
%    via resource-aware concurrency control in \S\ref{sec:discussion}.

\sreagent{} can be extended to support multiple concurrent writers
    by evolving A-Lock into a fine-grained, resource-aware concurrency control system.
The idea is to allow multiple mitigation agents 
    to execute transactions concurrently,
    provided their intended actions do not conflict over shared resources.
This can be achieved through a {\it Coordination Controller} that functions as a distributed lock manager.
Each mitigation agent first comes up with a plan $P_i$, 
    which is associated with a set of resources, $R(P_i)$, 
    containing the unique identifiers 
    of all system components its write actions will mutate. 
The controller maintains a global set of locked resources, $R_{\text{locked}}$, which is the union of resource sets for all currently executing plans, $R_{\text{locked}} = \bigcup_{j \in \text{executing}} R(P_j)$. A newly submitted plan $P_i$ is admitted for execution if and only if its resource set is disjoint from the global lock set: $\text{Admit}(P_i) \iff R(P_i) \cap R_{\text{locked}} = \emptyset$.

Upon admission, the plan's resources are atomically added to the global lock set 
    ($R_{\text{locked}} \leftarrow R_{\text{locked}} \cup R(P_i)$), effectively acquiring a distributed lock on them. 
The agent then executes its transaction under the original TNR semantics, guaranteeing that the severity metric does not increase: $\mu(s_{\text{post}}) \le \mu(s_{\text{pre}})$. 
Once the transaction completes (via commit or abort), its resources are released: $R_{\text{locked}} \leftarrow R_{\text{locked}} \setminus R(P_i)$. 
This mechanism upholds the ``Writer Exclusivity'' (A1) at the resource level, 
    ensuring that the safety properties of TNR hold for each concurrent transaction. 
Furthermore, this strict resource isolation can prevent read-after-write hazards, 
    as an agent is guaranteed a consistent view of any resource not explicitly locked by another transaction.
\vspace{-5pt}
\section{Implementation}
\label{sec:impl}
 \vspace{-5pt}

\begin{wrapfigure}{r}{0.275\textwidth}
    \centering
    \vspace{-30pt}
     \includegraphics[width=0.275\textwidth]{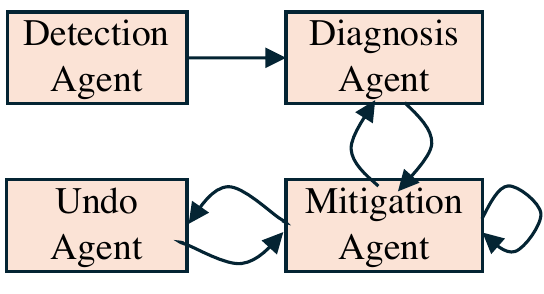}
     \vspace{-15pt}
     \caption{The state machine based control-flow logic.}
     \vspace{-5pt}
     \label{fig:state_machine}
 \end{wrapfigure}
\sreagent{} is implemented using the CrewAI multi-agent framework~\cite{crewai}.
The agents are orchestrated via state-machine based control-flow logic (Figure~\ref{fig:state_machine}).
\sreagent{} is an extensible framework.
The agents are loosely coupled and can be independently developed and customized.
We currently use off-the-shelf LLMs such as GPT and Llama models without fune-tuning.
One can also add new agents or replace existing agents.
The control plane can also be customized (e.g., using a different state machine)
    as it is decoupled from the data plane of the agents (each agent's data-flow logic).

\subsection{Realizing TNR in \sreagent{}}
\label{sec:impl:tnr}

In \sreagent{}, the assumptions A1--A3 are enforced as safety invariants with the following mechanisms.

% \begin{wrapfigure}{r}{0.27\textwidth}
%     \centering
%     % \vspace{-35pt}
%     \includegraphics[width=0.27\textwidth]{figures/state_machine.pdf}
%      \vspace{-15pt}
%     \caption{The state machine based control-flow logic.}
%     \vspace{-5pt}
%     \label{fig:state_machine}
% \end{wrapfigure}
{\bf Writer Exclusivity.} First, sandboxing-based confinement is implemented for each agent.
The detection and diagnosis agents are not allowed to issue any commands that may mutate system states (the 
    confinement rules are detailed in the appendix).
Second, the state machine ensures mutual exclusion when a writer agent (the mitigation agent or the undo agent) is
    scheduled to run, i.e., the sequences of the actions of the writer agents are strictly serialized.
We also instruct the agent to dry-run the commands (\code{kubectl} \code{--dry-run}) 
    whenever applicable to simulate command execution and identify errors without altering the system state.

{\bf Faithful Undo.} 
\sreagent{} ensures that every agent action has a corresponding undo operator; 
    otherwise, the action is not allowed.
\sreagent{} rejects destructive actions which cannot be recovered,
 or turns them into recoverable actions through the agent tools (e.g.,
    a file deletion is replaced by moving the file to a backup volume, which is recoverable).
A common pattern of the undo operator is system-state rollback. 
    Modern cloud platforms such as Kubernetes~\cite{Burns:2016}, Borg~\cite{borg}, 
        Twine~\cite{Tang:2020}, and ECS~\cite{ecs}
        are implemented based on {\it the state-reconciliation principle}~\cite{sun:hotos:21,sun:osdi:22},
        with declarative interfaces~\cite{gu:sosp:23,suresh:osdi:20}.
    Rollback in state-reconciliation systems can be implemented by recording the state changes and 
        later reconciling to the recorded states.

\begin{wrapfigure}{rt}{0.42\textwidth}
    \centering
    \vspace{-25pt}
    \includegraphics[width=0.42\textwidth]{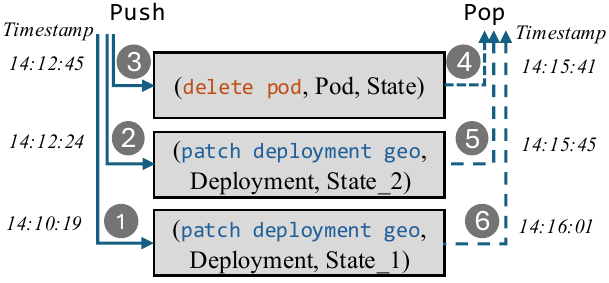}
    \caption{An example of the action stack used for reconciliation-based undo.}
    \label{fig:action-stack}
\end{wrapfigure}        
\sreagent{} implemented a stack-based rollback mechanism, as exemplified by Figure~\ref{fig:action-stack}.
The stack-based mechanism allows fine-grained rollback of the related system resources 
    (represented as state objects in Kubernetes) instead of the system-wide state.
Our current undo agent $\alpha_U$ mechanically walks over the stack and performs undo,
    and thus is arguably not intelligent.
The rationale of keeping $\alpha_U$ as an independent agent is to integrate more advanced 
    undo policies and mechanisms.
For example, we are exploring a learning-based rollback policy where the LLMs are given 
    the choice of undoing a subset of the actions in the stack (instead of always rolling back to the start state).

{\bf Bounded Risk Window.} The implementation is trivial. However, the window size 
    has an impact on mitigation effectiveness.
We set $K$ to 20 based on an empirical analysis (in appendix).

{\bf Limitations.} From an implementation perspective, realizing perfect undo for all conceivable state 
    changes in complex environments like cloud systems 
    remains a practical challenge (e.g., involving application-specific states and external interactions).
Certain operations may require more sophisticated compensation logic not covered by a simple $U(s_{\text{post}})=s_{\text{pre}}$. 
Fortunately, modern cloud-native systems are typically equipped with an operator program that follows 
    the state-reconciliation principle~\cite{Dobies2020,Ratis:2021}, which can be leveraged by \sreagent{}.
Moreover, our rule-based confinement may not be perfect to capture all destructive actions.
Recent work on advanced guardrails can help improve the confinement (see Related Work in \S\ref{sec:related}).

The principle of segmenting agent interventions into transactionally bounded stages, 
    each adhering to non-regression with respect to a measurable system health metric ($\mu$), 
    significantly de-risks autonomous operation. 
It transforms the problem from hoping an entire complex plan is perfect to verifying the safety of smaller, manageable steps.

\subsection{Other Notable Implementations}

{\bf Agent tools.} We develop tools that enable agents to interact with the environment
    using natural languages, following the Agent-Computer Interface (ACI) principles~\cite{yang2024swe}.
Specifically, we developed two sets of tools:
    (1) {\it observability tools} for agents to query different kinds of 
    observability data such as state objects (maintained by the cloud platform like Kubernetes), 
    system logs, distributed traces, performance metrics, etc;
We use LLMs to summarize the observability data instead of directly feeding a large volume 
    of data to the agents, as suggested by recent work~\cite{chen:eurosys:24}.
    (2) {\it command-line tools} that allow agents to execute commands and change system states
        using natural language,
    e.g., \code{NL2Kubectl} takes descriptions in natural language and 
        generates concrete \code{kubectl} commands.
    The confinement described in \S\ref{sec:impl:tnr} is implemented as the command-line tool wrappers.
\sreagent{}'s design supports hybrid operations
where writer agents can request human approval before execution,
with interceptable action stacks and tool interfaces.

{\bf Bootstrapping (Where shall the diagnosis agent start?)}
\label{sec:impl:bootstrapping}
The multi-agent design makes it easy to specialize agents with different roles.
We boost the diagnosis agent $\alpha_G$ with a bootstrapping technique that helps $\alpha_G$
    localize the failure region. 
In cloud systems, failures are propagated along request-response paths~\cite{Treynor:17,exchain},
    which is reflected by distributed traces~\cite{36356,kaldor2017canopy}.
\sreagent{} localizes the faults by constructing the call graph 
    and establishes an initial fault localization hypothesis to guide the diagnostic process.
Despite the simple idea, the fault-localization based 
    bootstrapping is important for large-scale systems and massive volumes of observability data.

{\bf Termination (When shall the agents stop?)}
\label{sec:term-oracle}
Agents need to correctly determine 
    whether the target failure is successfully resolved to stop its actions;
a perpetual agent could cause unintended effects,
    destabilize the system, or waste computing resources. 
\sreagent{} integrates a structured validation-and-termination approach.
After each diagnosis or mitigation procedure (a series of actions), 
    \sreagent{} assesses system health based on three kinds of oracles:
    (1) {\it alerts}: whether the alert that reports the target failure is cleared,
    (2) {\it user requests}: whether user requests can be successfully returned (e.g., no \code{5XX} or \code{4XX} HTTP responses),
    and (3) {\it system health}: whether system components (e.g., pods and volumes) are running in healthy states.
These oracles act as weak oracles individually, 
    each capturing a partial view of system health. 
\sreagent{} combines multiple oracles to form a stronger oracle. 
If all oracles pass, $\alpha_M$ concludes a successful mitigation and terminates.

 \vspace{-5pt}
\section{Evaluation}
\label{sec:eval}
\vspace{-5pt}

\begin{wrapfigure}{rt}{0.32\textwidth}
    \centering
    \vspace{-15pt}
    \includegraphics[width=0.325\textwidth]{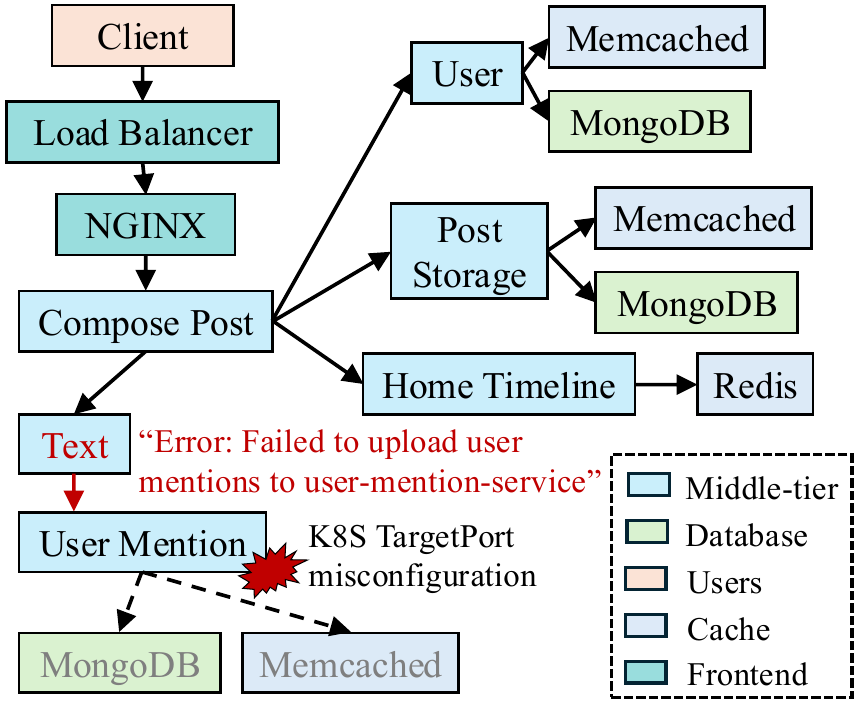}
    \caption{An example problem.}
    \vspace{-15pt}
    \label{fig:microservice-example}
\end{wrapfigure}

We evaluate \sreagent{} on two state-of-the-art benchmarks,
    AIOpsLab~\cite{chen2025aiopslab} and ITBench~\cite{jha2025itbench}.
Both benchmarks provide a live, arena-like environment,
    where AI agents are asked to resolve problems; each problem encodes 
    a failure in emulated cloud systems~\cite{deathstarbench,zhou2021fault,otelastronomy}, as exemplified by Figure~\ref{fig:microservice-example}.
We focus on \sreagent's ability to mitigate the failures (which has dependencies with detection and localization);
    we also report results of the other tasks, including detection, localization, and RCA.
Each benchmark takes \sreagent{} 15--20 machine hours to finish.

\paragraph{Baseline Agents.} We compare \sreagent{} with the reference agents offered by AIOpsLab and ITBench,
    referred to as AOL-agent and ITB-agent respectively.
For AIOpsLab, we also compare \sreagent{} with two other agents, 
    ReAct and Flash, included in the benchmark~\cite{chen2025aiopslab}.
We fuel these agents with different LLMs, including GPT-4o, GPT-4o-mini, and Llama 3.3.
The LLM temperature is set to 0 to make the results more deterministic.

We measure agentic systems using metrics:
    Success Rate (the percentage of problems being successfully solved), 
    average time (in seconds), the number of steps to solve a problem,
    and monetary cost in US dollars (calculated based on token consumptions).
We also profile the memory footprint 
    for one heavy problem
    across different agents in \sreagent.

{\bf Failure Mitigation Effectiveness.}
\label{sec:failure-mitigation-effectiveness}
As shown in \Cref{table:mitigation_results}, 
    \sreagent{} (GPT-4o) significantly outperforms state-of-the-art SRE agents on mitigation problems
    in both AIOpsLab and ITBench.
Concretely, it successfully solves 69.2\% (9/13) and 50.0\% (9/18)
    mitigation problems in AIOpsLab and ITBench, respectively.
This leads to improvements on the success rate by 1.5X and 5.4X 
    over the second best performing solution (AOL-agent and ITB-agent), respectively.
The advantage of \sreagent{} over the other SRE agents is consistent
    with different models, including both GPT-4o-mini and Llama 3.3.

We carefully inspect the problem-solving trajectories of the evaluated agents.
For the mitigation problems solved by \sreagent{} in AIOpsLab, 
    \sreagent{} correctly mitigated the triggering conditions or the root causes of the failures.
In eight ITBench problems, \sreagent{} exploits an observation that the injected faults do not persist,
    after the failing pods restarted (the fault injector cannot recognize the original pod); 
    thus, \sreagent{} tends to restart the failure pods 
    one by one and thus solves the problem.
Such a mitigation strategy may not work in reality, especially for failures caused by 
    persistent faults (e.g., misconfigurations and hardware defects).

\begin{table}[htbp]
    \centering
    \small
    % \vspace{-5pt}
%    \vspace{10pt}
    \caption{\textbf{Effectiveness of \sreagent{} in solving mitigation problems
        of AIOpsLab~\cite{chen2025aiopslab} and ITBench~\cite{jha2025itbench}.} 
        ITB-agent numbers are taken from~\cite{jha2025itbench} (``-'': unknown). 
        The ``\$'' column reports the average cost (in USD) of running the agent over all the mitigation tasks in each benchmark.}
    \label{table:mitigation_results}
    \begin{subtable}[t]{0.5\textwidth}
        \centering
        \caption{\bf AIOpsLab (13 Mitigation Problems)}
        \label{tab:metrics_mitigation}
        \setlength{\tabcolsep}{3.5pt}
        \begin{tabular}{lcccc}
          \toprule
          \textbf{Agent} & \textbf{Succ.} & \textbf{Time (s)} & \textbf{Steps} & \textbf{\$} \\
          \midrule
          ReAct (4o)   & \cellcolor{mellowgreen!23} 23.1\% & 46.0 & 23.0 & 0.112 \\
          Flash (4o)   & \cellcolor{mellowgreen!38} 38.5\% & 154.0 & 23.1 & 0.150 \\
          \midrule
          AOL-agent (4o)    & \cellcolor{mellowgreen!82} 46.2\% & 223.3 & 21.7 & 0.206 \\
          AOL-agent (mini)  & \cellcolor{mellowgreen!7} 7.7\% & 58.9 & 22.7 & 0.003 \\
          AOL-agent (llama) & \cellcolor{mellowgreen!15} 15.4\% & 98.2 & 13.0 & 0.037 \\
          \midrule
          \sreagent{} (4o) & \cellcolor{mellowgreen!100} 69.2\% & 811.9 & 46.3 & 0.877 \\
          \sreagent{} (mini) & 	\cellcolor{mellowgreen!23} 23.1\% & 3557.9 & 125.7 & 0.036 \\
          \sreagent{} (llama) &  \cellcolor{mellowgreen!23} 23.1\% & 1486.9 & 71.8 & 0.360 \\
          \bottomrule
          \end{tabular}
    \end{subtable}
    \hfill
    \begin{subtable}[t]{0.475\textwidth}
        \centering
        \caption{\bf ITBench (18 Mitigation Problems)}
        \label{tab:metrics_itbench_lite}
        \setlength{\tabcolsep}{3.5pt}
        \begin{tabular}{lcccc}
        \toprule
        \textbf{Agent} & \textbf{Succ.} & \textbf{Time (s)} & \textbf{Steps} & \textbf{\$} \\
        \midrule
        ITB-agent (4o) & \cellcolor{mellowgreen!30} 9.2\% & 251.7 & - & - \\
        ITB-agent (llama) & \cellcolor{mellowgreen!25} 5.7\% & 440.8 & - & - \\
        \midrule
        \sreagent{} (4o) & \cellcolor{mellowgreen!100} 50.0\% & 1720.8 & 115.7 & 6.11 \\
        \sreagent{} (mini) & \cellcolor{mellowgreen!40} 19.4\% & 3874.9 & 468.9 & 9.38 \\
        \sreagent{} (llama) & \cellcolor{mellowgreen!62} 28.0\% & 2566.6 & 160.3 & 0.76 \\
        \bottomrule
        \end{tabular}
        {\scriptsize
        \begin{itemize}
        \item ``4o'' refers to GPT-4o (gpt-4o-2024-08-06) \\[-5pt]
        \item ``mini'' refers to GPT-4o-mini (gpt-4o-mini-2024-07-18)\\[-5pt]
        \item ``llama'' refers to Llama 3.3 (llama-3-3-70b-instruct)\\[-5pt]
        \end{itemize}
        }
      \end{subtable}
    %   \vspace{-5pt}
  \end{table}

\sreagent{} overall takes longer time to solve problems;
for the hard problems, \sreagent{} retries multiple times to solve them---{\it the retry is enabled 
    by the Transactional No-Regression safety guarantee.}
In comparison, the other agents do not have the safe retry capability.
In fact, they provide no safety guarantee over their mitigation actions---for the failures 
    they fail to mitigate, these agents can leave the system in worse states
    than the original error state.
For the same reason, \sreagent{} spends more in dollar amount, but the amount 
    is significantly cheaper than human cost.

{\bf Effectiveness of TNR-based Undo-and-Retry.}
\label{sec:undo-and-retry}
The TNR safety is crucial to \sreagent{}'s effectiveness in solving 
    complex mitigation problems.
Table~\ref{table:ablation_mitigation} shows this point through an ablation study
    of \sreagent{} (GPT-4o)
    (more detailed ablation analysis can be found in the appendix).
We compare \sreagent{} (GPT-4o) with two variants:
(1) {\it No retry:} Only attempting one mitigation path, and 
(2) {\it Na\"{i}ve retry w/o undo:} exploring a new mitigation path if the previous
        mitigation path was not successful; however, the new exploration is conducted 
        directly from the ending state of the last attempt, instead of rolling back 
        to the original state as in \sreagent{}.

We see that (1) the ability of retrying a different mitigation plan is important---only 
    in two problems, the first mitigation plan generated by \sreagent{} successfully mitigates the failure,
and (2) the ability to undo and rollback the system state is critical---new mitigation from an error state 
    left by a failed attempt can hardly succeed (only one new problem is solved in this way).
Intuitively, failed mitigations further worsen the system states, making the system harder and harder to save.
TNR prevents such failure patterns and thus is essential to mitigation.
Figure~\ref{fig:retry_cdf} shows the probability density function of the number of retries (per problem)
    done by the \sreagent{} (GPT-4o) in AIOpsLab mitigation problems.
\sreagent{} sets a limit of nine retries.
In 80+\% of the problems, \sreagent{} retries at least once;
    in 30+\% of the problems, \sreagent{} retries no less than five times.
    
    \begin{minipage}{\textwidth}
        \small
        \centering
        \vspace{10pt}
        \begin{minipage}[t]{0.625\textwidth}
            \centering
            \captionof{table}{{\bf Ablation analysis on TNR-enabled undo-and-retry 
                of \sreagent{} (GPT-4o)} using the 13 mitigation problems in AIOpsLab,
                which shows the importance of TNR safety.}
            \vspace{0pt}
            \begin{tabular}{lccc}
                \toprule
                \textbf{Ablation} & \textbf{Succ. Rate} & \textbf{Time (s)} & \textbf{Cost (\$)} \\
                \midrule
                \sreagent{} (4o) & 69.2\% & 811.9 & 0.877 \\
                $-$ No retry & 15.4\% & 72.6 & 0.163 \\
                $-$ Na\"{i}ve retry w/o undo & 23.1\% & 1221.5 & 0.929 \\
                \bottomrule
            \end{tabular}
            \label{table:ablation_mitigation}
        \end{minipage}%
        \hfill
        \begin{minipage}[t]{0.325\textwidth}
            \centering
            \vspace{0pt}
            \includegraphics[width=\linewidth]{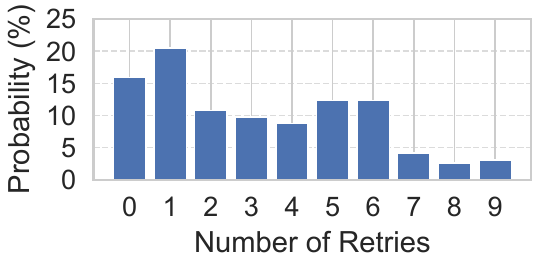}
            \captionof{figure}{Probability density of the retry times per problem.}
            \label{fig:retry_cdf}
        \end{minipage}
        \vspace{10pt}
    \end{minipage}

{\bf Effectiveness on Detection, Localization, and Root Cause Analysis (RCA).} We 
    also evaluated the effectiveness of \sreagent{} on other SRE problems,
    in comparison to other SRE agents.
\Cref{table:agent_performance_by_tasks} shows the results for problems 
    in AIOpsLab.
Note that the number of problems in the categories are different.
\sreagent{} outperforms the reference agents in 
    both detection and localization problems.
Specifically, \sreagent{} with Llama and GPT-4o achieves 
    over 90\% of success rate for detection.
Localization and RCA problems, which require the agents
    to pinpoint the root-cause components (e.g., the pod)
    and the fault type (e.g., misconfiguration or bug),
    present greater challenges to agentic solutions;
\sreagent{} (GPT-4o) has the highest success rate
    for localization problems and is second highest
    for RCA problems. Note that RCA is not a prerequisite of failure mitigation (see \S\ref{sec:background}).

\begin{table}[htbp]
    \centering
    \vspace{-7.5pt}
    \caption{\textbf{Effectiveness of \sreagent{} in solving other SRE problems in AIOpsLab.}
      The problems in AIOpsLab are independent; there are in total 86 problems of four SRE types (\S\ref{sec:background}).}
    \footnotesize
      \setlength{\tabcolsep}{2.5pt}
      \vspace{3.5pt}
      \begin{tabular}{lccccccccccc}
      \toprule
                     & \multicolumn{3}{c}{\bf Detection (32 Problems)} & & \multicolumn{3}{c}{\bf Localization (28 Problems)} & & \multicolumn{3}{c}{\bf RCA (26 Problems)} \\
      \cline{2-4}
      \cline{6-8}
      \cline{10-12}
      \textbf{Agent\ } & \textbf{Succ.} & \textbf{Time (s)} & \textbf{\$} & & \textbf{Succ.} & \textbf{Time (s)} & \textbf{\$} & & \textbf{Succ.} & \textbf{Time (s)} & \textbf{\$}\\
      \midrule
      ReAct (4o) & \cellcolor{mellowgreen!87} 87.5\% & 33.2 & 0.086 & & \cellcolor{mellowgreen!52} 26.8\% & 59.6 & 0.328 & & \cellcolor{mellowgreen!23} 23.1\% & 28.5 & 0.065 \\
      Flash (4o) & \cellcolor{mellowgreen!28} 59.4\% & 30.7 & 0.013 & & \cellcolor{mellowgreen!78} 39.3\% & 165.0 & 0.190 & & \cellcolor{mellowgreen!65} 26.9\% & 30.5 & 0.019 \\
      \midrule
      AOL-agent (4o) & \cellcolor{mellowgreen!31} 62.5\% & 14.4 & 0.061 & & \cellcolor{mellowgreen!92} 46.9\% & 34.8 & 0.083 & & \cellcolor{mellowgreen!100} 38.5\% & 12.3 & 0.061 \\ 
      AOL-agent (mini) & \cellcolor{mellowgreen!12} 25.0\% & 43.0 & 0.002 & & \cellcolor{mellowgreen!9} 9.5\% & 34.1 & 0.001 & & \cellcolor{mellowgreen!19} 7.7\% & 57.7 & 0.003 \\
      AOL-agent (llama) & \cellcolor{mellowgreen!42} 84.4\% & 19.8 & 0.019 & & \cellcolor{mellowgreen!64} 32.1\% & 40.5 & 0.018 & & \cellcolor{mellowgreen!85} 30.8\% & 13.8 & 0.014 \\
      \midrule
      \sreagent{} (4o) & \cellcolor{mellowgreen!100} 90.6\% & 48.4 & 0.118 & & \cellcolor{mellowgreen!100} 51.2\% & 65.3 & 0.126 & & \cellcolor{mellowgreen!95} 34.6\% & 39.6 & 0.068 \\
      \sreagent{} (mini) & \cellcolor{mellowgreen!78} 78.1\% & 34.4 & 0.010 & & \cellcolor{mellowgreen!50} 25.0\% & 37.2 & 0.013 & & \cellcolor{mellowgreen!90} 30.8\% & 279.0 & 0.007 \\
      \sreagent{} (llama) & \cellcolor{mellowgreen!93} 93.8\% & 50.0 & 0.111 & & \cellcolor{mellowgreen!72} 36.3\% & 90.5 & 0.112 & & \cellcolor{mellowgreen!65} 26.9\% & 60.2 & 0.095 \\
      \bottomrule
      \end{tabular}
      \label{table:agent_performance_by_tasks}
  \end{table}

Note that the relatively low RCA success rate across other tasks for \sreagent{} (and all the other agents) 
    is largely attributed to the ambiguous evaluation metrics used by AIOpsLab.
AIOpsLab measures RCA success by matching the agent's output against
    pre-defined category labels; however, some labels are ambiguous and not mutually exclusive.
For example, for a port misconfiguration problem, AIOpsLab defines the root cause
    to be ``Virtualization'' (as the layer) and ``Misconfiguration'' (as the fault type).
\sreagent{} answers the root cause to be ``Application'' (as the layer)
    and ``Network/Storage Issue'' (as the fault type).
Arguably, \sreagent{}'s answer is more precise in describing the root cause---a port
    is more of an application-level issue than ``Virtualization'',
    since there is no traditional hypervisor-like virtualization layer in AIOpsLab.
However, \sreagent{} is considered incorrect for this problem in AIOpsLab due to such label mismatch.
As ongoing work, we are actively investigating new metrics designs to improve the benchmarks for SRE agents.

% Our agents have no cold-start delay as they directly use existing LLMs
%     without maintaining long-term memory.
% \yf{
% While \sreagent{} takes longer on harder problems (Table~\ref{table:agent_performance_by_tasks}),
%    due to TNR-enabled retries,
%    this is a key advantage rather than a limitation---other agents
%    cannot safely retry and often leave systems in worse states.
% We envision several acceleration strategies:
%     (1) caching solutions for similar failure types,
%     (2) using smaller models for specific agents (e.g., Llama3.3-70B for Undo),
%     and (3) reasoning about mitigation strategy efficiency
%     (e.g., cheapest-first strategies).

% Through manual inspection of all \sreagent{} RCA answers,
%     we found the true success rate of \sreagent{} (GPT-4o) to be 50\% (13/26),
%     which was confirmed by the AIOpsLab authors who acknowledged this evaluation limitation.

{\bf Memory usage.}
To understand the memory usage of \sreagent,  
    we profile \sreagent{} running a heavy mitigation problem
    (\code{misconfig_app_hotel_res-mitigation-1} in AIOpsLab),
    which takes 13 steps to resolve.
% The memory footprints are shown in Table~\ref{tab:memory}.
The dominant memory cost (over 530 MB) comes from loading the agent runtime
    and its dependencies (e.g., CrewAI and benchmark tooling).
The Detection agent uses the most memory as it collects and analyzes raw telemetry data
    (e.g., using \texttt{pandas} for trace analysis).
The Mitigation agent consumes the largest context memory (31.03 KB) 
    as it accumulates trajectories across steps, 
    whereas the Undo agent uses the least (5.79 KB) 
    due to its simple rollback operations.

\section{Related Work}
\label{sec:related}
\vspace{-3.5pt}

{\bf AI and Agentic Research for SRE.}
Applying AI/ML techniques to solve SRE problems 
    has a long history of research,
    as AI/ML has the potential to accommodate the large scale 
    of modern systems and infrastructures (e.g., the clouds)~\cite{zhou2021fault,miao2010generative},
    as well as massive volumes of observability data~\cite{kaldor2017canopy}.
Specialized AI/ML techniques are designed for concrete SRE tasks,
    such as failure detection~\cite{poth:15,zhang:14,palatin:06},
    triage~\cite{shao2008efficient,shao2008easyticket}, and diagnosis~\cite{ikram2022root,gan2021sage,li2022causal,mickens:07}.
Recently, generative AI models, as well as agentic approaches, 
    are increasingly used to further generalize and automate these SRE tasks~\cite{ahmed2023recommending,chen:eurosys:24,yu2024monitorassistant,jiang2024xpert}.
However, the aforementioned efforts focus on building tools to assist human engineering (see \S\ref{sec:background}).
Differently, we aim for autonomous SRE. A differential feature is {\it failure mitigation}.
Prior work mostly focuses on helping human engineers to detect, understand the failure or recommend potential mitigation documents, 
    where mitigation is the main goal of \sreagent{}

{\bf Safety of Agentic AI Systems.} 
Safety is an emerging concern of agentic AI systems~\cite{miehling2025agenticaineedssystems,shinn2023reflexionlanguageagentsverbal,dong2024safeguardinglargelanguagemodels,hua2024trustagent}.
    Prior work focuses on implementing safety guardrails~\cite{nunez2024autosafecodermultiagentframeworksecuring, inan2023llamaguardllmbasedinputoutput, guardrailsai, wang2025agentspeccustomizableruntimeenforcement}, 
    which aim to prevent agents from producing or acting upon harmful inputs and outputs, often through prompt-level or static analysis.
    However, such preventative approaches are limited in environments like cloud systems, 
        where the system state evolves over time and side effects emerge dynamically only after actions are executed.
    For example, AutoSafeCoder~\cite{nunez2024autosafecodermultiagentframeworksecuring} mitigates code vulnerabilities via static checks, 
    but such methods cannot anticipate runtime failures or cascading effects that depend on temporal system dynamics.
In addition, few works provide guarantees on recovery.
For example, TrustAgent~\cite{hua2024trustagent} proposes the Agent-Constitution framework that address agent safety.
However, it does not provide guarantees on the safety of the agent action or any recovery from unsafe actions.
For SRE, this could allow failures to regress into  a worse situation due to unsafe agent actions.

\sreagent{} addresses this challenge by enabling safe exploration with Transactional Non-Regression (TNR).
    TNR allows agents to explore multiple action sequences safely 
    and iteratively, giving the agent more opportunities to mitigate failures. 
Concurrent work has also started integrating transactional primitives into LLM-based frameworks~\cite{chang2025sagallm}.
    \sreagent{} extends this direction by enabling rollback of both the agent and the system state, combined with a continuous severity metric to assess action outcomes.
    This dynamic and fine-grained safety mechanism is critical for real-world deployment of autonomous agents in evolving operational contexts like SRE.

{\bf Multi-agent Systems.} 
Multi-agent has become a common design pattern, with 
    the development of many LLM-based multi-agent frameworks~\cite{crewai,li2023camel,autogen,hong2024metagpt,openaiswarm}
    and use cases~\cite{qian2024chatdev,xu2024ai,zhou2024large,zhao-etal-2024-longagent}.
Recent work proposes multi-agent conversations~\cite{autogen}
    and debates~\cite{du2023improving,liang2023encouraging}
    that encourage diverse thinking among agents.
However, our experience shows that conversations and debates 
    are not suitable for SRE tasks which require safety reasoning
    and timeliness of problem solving.
Hence, \sreagent{} uses a deterministic state machine to coordinate the agents.
Recent work also reports common failure modes of multi-agent systems~\cite{zhang2025agentcausestaskfailures,cemri2025multiagentllmsystemsfail}.
We detail how \sreagent{} addresses them in the appendix.

\section{Concluding Remarks}
\vspace{-3.5pt}
With generative AI and agentic technologies, autonomous SRE for modern clouds may no longer 
    be a pipedream.
While we can expect new AI models and fine-tuning to further improve 
    problem-solving abilities of our agents,
    we assert that safety is the key challenge for agentic SRE systems
    to operate live production clouds.
We also assert that agent safety has to be rigorously defined and enforced as a first-class system design principle.
This paper makes a first attempt to formalize TNR as a useful safety specification of agentic SRE
    and realize it in our multi-agent SRE system.
We envision and hope the work to inspire stronger safety properties 
    and system-level support for agentic SRE systems.

\section*{Acknowledgment}

We thank the anonymous reviewers for their insightful comments.
The work is supported in part by 
    an IBM-Illinois Discovery Accelerator Institute (IIDAI) grant
    and NSF CNS-2145295.

\clearpage
\bibliographystyle{acm}
\bibliography{ref}

\newpage
\section*{NeurIPS Paper Checklist}

{\bf 1. Claims}

{\bf Question:} Do the main claims made in the abstract and introduction accurately reflect the paper's contributions and scope?

{\bf Answer:} \answerYes{} % Replace by \answerYes{}, \answerNo{}, or \answerNA{}.

{\bf Justification:} We list our main contributions in the end of the Introduction,
    and the contributions are analyzed in Section~\ref{sec:sreagent} and evaluated in Section~\ref{sec:eval}.

{\bf Guidelines:}

    \begin{packed_itemize}
        \item The answer NA means that the abstract and introduction do not include the claims made in the paper.
        \item The abstract and/or introduction should clearly state the claims made, including the contributions made in the paper and important assumptions and limitations. A No or NA answer to this question will not be perceived well by the reviewers. 
        \item The claims made should match theoretical and experimental results, and reflect how much the results can be expected to generalize to other settings. 
        \item It is fine to include aspirational goals as motivation as long as it is clear that these goals are not attained by the paper. 
    \end{packed_itemize}

{\bf 2. Limitations}

{\bf Question:} Does the paper discuss the limitations of the work performed by the authors?

{\bf Answer:} \answerYes{} % Replace by \answerYes{}, \answerNo{}, or \answerNA{}.

{\bf Justification:} Limitations are discussed in context (see Section~\ref{sec:model:discussion} and Section~\ref{sec:impl}).

{\bf Guidelines:}
    \begin{packed_itemize}
        \item The answer NA means that the paper has no limitation while the answer No means that the paper has limitations, but those are not discussed in the paper. 
        \item The authors are encouraged to create a separate "Limitations" section in their paper.
        \item The paper should point out any strong assumptions and how robust the results are to violations of these assumptions (e.g., independence assumptions, noiseless settings, model well-specification, asymptotic approximations only holding locally). The authors should reflect on how these assumptions might be violated in practice and what the implications would be.
        \item The authors should reflect on the scope of the claims made, e.g., if the approach was only tested on a few datasets or with a few runs. In general, empirical results often depend on implicit assumptions, which should be articulated.
        \item The authors should reflect on the factors that influence the performance of the approach. For example, a facial recognition algorithm may perform poorly when image resolution is low or images are taken in low lighting. Or a speech-to-text system might not be used reliably to provide closed captions for online lectures because it fails to handle technical jargon.
        \item The authors should discuss the computational efficiency of the proposed algorithms and how they scale with dataset size.
        \item If applicable, the authors should discuss possible limitations of their approach to address problems of privacy and fairness.
        \item While the authors might fear that complete honesty about limitations might be used by reviewers as grounds for rejection, a worse outcome might be that reviewers discover limitations that aren't acknowledged in the paper. The authors should use their best judgment and recognize that individual actions in favor of transparency play an important role in developing norms that preserve the integrity of the community. Reviewers will be specifically instructed to not penalize honesty concerning limitations.
    \end{packed_itemize}

{\bf 3. Theory assumptions and proofs}

{\bf Question:} For each theoretical result, does the paper provide the full set of assumptions and a complete (and correct) proof?

{\bf Answer:} \answerYes{} % Replace by \answerYes{}, \answerNo{}, or \answerNA{}.

{\bf Justification:} We state our assumptions in Section~\ref{subsec:assumptions}
 and write down our proofs in Section~\ref{subsec:tnr}.

{\bf Guidelines:}
    \begin{packed_itemize}
        \item The answer NA means that the paper does not include theoretical results. 
        \item All the theorems, formulas, and proofs in the paper should be numbered and cross-referenced.
        \item All assumptions should be clearly stated or referenced in the statement of any theorems.
        \item The proofs can either appear in the main paper or the supplemental material, but if they appear in the supplemental material, the authors are encouraged to provide a short proof sketch to provide intuition. 
        \item Inversely, any informal proof provided in the core of the paper should be complemented by formal proofs provided in appendix or supplemental material.
        \item Theorems and Lemmas that the proof relies upon should be properly referenced. 
    \end{packed_itemize}

{\bf 4. Experimental result reproducibility}

{\bf Question:} Does the paper fully disclose all the information needed to reproduce the main experimental results 
of the paper to the extent that it affects the main claims and/or conclusions of the paper 
(regardless of whether the code and data are provided or not)?

{\bf Answer:} \answerYes{} % Replace by \answerYes{}, \answerNo{}, or \answerNA{}.

{\bf Justification:} Section~\ref{sec:impl} discusses our implementation. 
The artifacts and instructions can be found at \url{https://github.com/xlab-uiuc/stratus}.
% We will release our code on GitHub with a replication package that describes steps to reproduce the results.

{\bf Guidelines:}
    \begin{packed_itemize}
        \item The answer NA means that the paper does not include experiments.
        \item If the paper includes experiments, a No answer to this question will not be perceived well by the reviewers: Making the paper reproducible is important, regardless of whether the code and data are provided or not.
        \item If the contribution is a dataset and/or model, the authors should describe the steps taken to make their results reproducible or verifiable. 
        \item Depending on the contribution, reproducibility can be accomplished in various ways. For example, if the contribution is a novel architecture, describing the architecture fully might suffice, or if the contribution is a specific model and empirical evaluation, it may be necessary to either make it possible for others to replicate the model with the same dataset, or provide access to the model. In general. releasing code and data is often one good way to accomplish this, but reproducibility can also be provided via detailed instructions for how to replicate the results, access to a hosted model (e.g., in the case of a large language model), releasing of a model checkpoint, or other means that are appropriate to the research performed.
        \item While NeurIPS does not require releasing code, the conference does require all submissions to provide some reasonable avenue for reproducibility, which may depend on the nature of the contribution. For example
        \begin{enumerate}
            \item If the contribution is primarily a new algorithm, the paper should make it clear how to reproduce that algorithm.
            \item If the contribution is primarily a new model architecture, the paper should describe the architecture clearly and fully.
            \item If the contribution is a new model (e.g., a large language model), then there should either be a way to access this model for reproducing the results or a way to reproduce the model (e.g., with an open-source dataset or instructions for how to construct the dataset).
            \item We recognize that reproducibility may be tricky in some cases, in which case authors are welcome to describe the particular way they provide for reproducibility. In the case of closed-source models, it may be that access to the model is limited in some way (e.g., to registered users), but it should be possible for other researchers to have some path to reproducing or verifying the results.
        \end{enumerate}
    \end{packed_itemize}

{\bf 5. Open access to data and code}

{\bf Question:} Does the paper provide open access to the data and code, with sufficient instructions to faithfully reproduce the main experimental results, as described in supplemental material?

{\bf Answer:} \answerYes{} % Replace by \answerYes{}, \answerNo{}, or \answerNA{}.

{\bf Justification:} Section~\ref{sec:impl} discusses our implementation. 
The artifacts and instructions can be found at \url{https://github.com/xlab-uiuc/stratus}.
% We will release our code on GitHub with a replication package that describes steps to reproduce the results.

{\bf Guidelines:}
    \begin{packed_itemize}
        \item The answer NA means that paper does not include experiments requiring code.
        \item Please see the NeurIPS code and data submission guidelines (\url{https://nips.cc/public/guides/CodeSubmissionPolicy}) for more details.
        \item While we encourage the release of code and data, we understand that this might not be possible, so “No” is an acceptable answer. Papers cannot be rejected simply for not including code, unless this is central to the contribution (e.g., for a new open-source benchmark).
        \item The instructions should contain the exact command and environment needed to run to reproduce the results. See the NeurIPS code and data submission guidelines (\url{https://nips.cc/public/guides/CodeSubmissionPolicy}) for more details.
        \item The authors should provide instructions on data access and preparation, including how to access the raw data, preprocessed data, intermediate data, and generated data, etc.
        \item The authors should provide scripts to reproduce all experimental results for the new proposed method and baselines. If only a subset of experiments are reproducible, they should state which ones are omitted from the script and why.
        \item At submission time, to preserve anonymity, the authors should release anonymized versions (if applicable).
        \item Providing as much information as possible in supplemental material (appended to the paper) is recommended, but including URLs to data and code is permitted.
    \end{packed_itemize}

{\bf 6. Experimental setting/details}

{\bf Question:} Does the paper specify all the training and test details (e.g., data splits, hyperparameters, how they were chosen, type of optimizer, etc.) necessary to understand the results?

{\bf Answer:} \answerYes{} % Replace by \answerYes{}, \answerNo{}, or \answerNA{}.

{\bf Justification:} We explain our experimental setting and details in Section~\ref{sec:eval} and appendix.

{\bf Guidelines:}
    \begin{packed_itemize}
        \item The answer NA means that the paper does not include experiments.
        \item The experimental setting should be presented in the core of the paper to a level of detail that is necessary to appreciate the results and make sense of them.
        \item The full details can be provided either with the code, in appendix, or as supplemental material.
    \end{packed_itemize}

{\bf 7. Experiment statistical significance}

{\bf Question:} Does the paper report error bars suitably and correctly defined or other appropriate information about the statistical significance of the experiments?

{\bf Answer:} \answerYes{} % Replace by \answerYes{}, \answerNo{}, or \answerNA{}.

{\bf Justification:} The reported results are statistically significant. 
    The experiments are run multiple times and the median numbers are reported in the tables.

{\bf Guidelines:}
    \begin{packed_itemize}
        \item The answer NA means that the paper does not include experiments.
        \item The authors should answer "Yes" if the results are accompanied by error bars, confidence intervals, or statistical significance tests, at least for the experiments that support the main claims of the paper.
        \item The factors of variability that the error bars are capturing should be clearly stated (for example, train/test split, initialization, random drawing of some parameter, or overall run with given experimental conditions).
        \item The method for calculating the error bars should be explained (closed form formula, call to a library function, bootstrap, etc.)
        \item The assumptions made should be given (e.g., Normally distributed errors).
        \item It should be clear whether the error bar is the standard deviation or the standard error of the mean.
        \item It is OK to report 1-sigma error bars, but one should state it. The authors should preferably report a 2-sigma error bar than state that they have a 96\% CI, if the hypothesis of Normality of errors is not verified.
        \item For asymmetric distributions, the authors should be careful not to show in tables or figures symmetric error bars that would yield results that are out of range (e.g. negative error rates).
        \item If error bars are reported in tables or plots, The authors should explain in the text how they were calculated and reference the corresponding figures or tables in the text.
    \end{packed_itemize}

{\bf 8. Experiments compute resources}

{\bf Question:} For each experiment, does the paper provide sufficient information on the computer resources (type of compute workers, memory, time of execution) needed to reproduce the experiments?

{\bf Answer:} \answerYes{} % Replace by \answerYes{}, \answerNo{}, or \answerNA{}.

{\bf Justification:} We document the compute resources used in the appendix.
    We list the cost of using LLMs in our evaluation section. {}

{\bf Guidelines:}
    \begin{packed_itemize}
        \item The answer NA means that the paper does not include experiments.
        \item The paper should indicate the type of compute workers CPU or GPU, internal cluster, or cloud provider, including relevant memory and storage.
        \item The paper should provide the amount of compute required for each of the individual experimental runs as well as estimate the total compute. 
        \item The paper should disclose whether the full research project required more compute than the experiments reported in the paper (e.g., preliminary or failed experiments that didn't make it into the paper). 
    \end{packed_itemize}
    
{\bf 9. Code of ethics}

{\bf Question:} Does the research conducted in the paper conform, in every respect, with the NeurIPS Code of Ethics \url{https://neurips.cc/public/EthicsGuidelines}?

{\bf Answer:} \answerYes{} % Replace by \answerYes{}, \answerNo{}, or \answerNA{}.

{\bf Justification:} All experiments are conducted on publicly available benchmarks. 
    We do not anticipate any negative social impacts. 
    Our agent is designed specifically for Site Reliability Engineering.{}

{\bf Guidelines:}
    \begin{packed_itemize}
        \item The answer NA means that the authors have not reviewed the NeurIPS Code of Ethics.
        \item If the authors answer No, they should explain the special circumstances that require a deviation from the Code of Ethics.
        \item The authors should make sure to preserve anonymity (e.g., if there is a special consideration due to laws or regulations in their jurisdiction).
    \end{packed_itemize}

{\bf 10. Broader impacts}

{\bf Question:} Does the paper discuss both potential positive societal impacts and negative societal impacts of the work performed?

{\bf Answer:} \answerYes{} % Replace by \answerYes{}, \answerNo{}, or \answerNA{}.

{\bf Justification:} We discuss broader societal impacts in the appendix.

{\bf Guidelines:}
    \begin{packed_itemize}
        \item The answer NA means that there is no societal impact of the work performed.
        \item If the authors answer NA or No, they should explain why their work has no societal impact or why the paper does not address societal impact.
        \item Examples of negative societal impacts include potential malicious or unintended uses (e.g., disinformation, generating fake profiles, surveillance), fairness considerations (e.g., deployment of technologies that could make decisions that unfairly impact specific groups), privacy considerations, and security considerations.
        \item The conference expects that many papers will be foundational research and not tied to particular applications, let alone deployments. However, if there is a direct path to any negative applications, the authors should point it out. For example, it is legitimate to point out that an improvement in the quality of generative models could be used to generate deepfakes for disinformation. On the other hand, it is not needed to point out that a generic algorithm for optimizing neural networks could enable people to train models that generate Deepfakes faster.
        \item The authors should consider possible harms that could arise when the technology is being used as intended and functioning correctly, harms that could arise when the technology is being used as intended but gives incorrect results, and harms following from (intentional or unintentional) misuse of the technology.
        \item If there are negative societal impacts, the authors could also discuss possible mitigation strategies (e.g., gated release of models, providing defenses in addition to attacks, mechanisms for monitoring misuse, mechanisms to monitor how a system learns from feedback over time, improving the efficiency and accessibility of ML).
    \end{packed_itemize}
    
{\bf 11. Safeguards}

{\bf Question:} Does the paper describe safeguards that have been put in place for responsible release of data or models that have a high risk for misuse (e.g., pretrained language models, image generators, or scraped datasets)?

{\bf Answer:} \answerNA{} % Replace by \answerYes{}, \answerNo{}, or \answerNA{}.

{\bf Justification:} We do not release any data or models, so there is no potential for misuse.{}

{\bf Guidelines:}
    \begin{packed_itemize}
        \item The answer NA means that the paper poses no such risks.
        \item Released models that have a high risk for misuse or dual-use should be released with necessary safeguards to allow for controlled use of the model, for example by requiring that users adhere to usage guidelines or restrictions to access the model or implementing safety filters. 
        \item Datasets that have been scraped from the Internet could pose safety risks. The authors should describe how they avoided releasing unsafe images.
        \item We recognize that providing effective safeguards is challenging, and many papers do not require this, but we encourage authors to take this into account and make a best faith effort.
    \end{packed_itemize}

{\bf 12. Licenses for existing assets}

{\bf Question:} Are the creators or original owners of assets (e.g., code, data, models), used in the paper, properly credited and are the license and terms of use explicitly mentioned and properly respected?

{\bf Answer:} \answerYes{} % Replace by \answerYes{}, \answerNo{}, or \answerNA{}.

{\bf Justification:} The foundation models used in our experiments are cited properly.
    Only open-source benchmark suites are used.

{\bf Guidelines:}
    \begin{packed_itemize}
        \item The answer NA means that the paper does not use existing assets.
        \item The authors should cite the original paper that produced the code package or dataset.
        \item The authors should state which version of the asset is used and, if possible, include a URL.
        \item The name of the license (e.g., CC-BY 4.0) should be included for each asset.
        \item For scraped data from a particular source (e.g., website), the copyright and terms of service of that source should be provided.
        \item If assets are released, the license, copyright information, and terms of use in the package should be provided. For popular datasets, \url{paperswithcode.com/datasets} has curated licenses for some datasets. Their licensing guide can help determine the license of a dataset.
        \item For existing datasets that are re-packaged, both the original license and the license of the derived asset (if it has changed) should be provided.
        \item If this information is not available online, the authors are encouraged to reach out to the asset's creators.
    \end{packed_itemize}

{\bf 13. New assets}

{\bf Question:} Are new assets introduced in the paper well documented and is the documentation provided alongside the assets?

{\bf Answer:} \answerYes{} % Replace by \answerYes{}, \answerNo{}, or \answerNA{}.

{\bf Justification:} Our only new asset is our code, can be found at \url{https://github.com/xlab-uiuc/stratus}. 
    We also provide scripts and a README file containing documentation in the repository.{}

{\bf Guidelines:}
    \begin{packed_itemize}
        \item The answer NA means that the paper does not release new assets.
        \item Researchers should communicate the details of the dataset/code/model as part of their submissions via structured templates. This includes details about training, license, limitations, etc. 
        \item The paper should discuss whether and how consent was obtained from people whose asset is used.
        \item At submission time, remember to anonymize your assets (if applicable). You can either create an anonymized URL or include an anonymized zip file.
    \end{packed_itemize}

{\bf 14. Crowdsourcing and research with human subjects}

{\bf Question:} For crowdsourcing experiments and research with human subjects, does the paper include the full text of instructions given to participants and screenshots, if applicable, as well as details about compensation (if any)? 

{\bf Answer:} \answerNA{} % Replace by \answerYes{}, \answerNo{}, or \answerNA{}.

{\bf Justification:} This paper does not involve crowdsourcing or research with any human subjects.{}

{\bf Guidelines:}
    \begin{packed_itemize}
        \item The answer NA means that the paper does not involve crowdsourcing nor research with human subjects.
        \item Including this information in the supplemental material is fine, but if the main contribution of the paper involves human subjects, then as much detail as possible should be included in the main paper. 
        \item According to the NeurIPS Code of Ethics, workers involved in data collection, curation, or other labor should be paid at least the minimum wage in the country of the data collector. 
    \end{packed_itemize}

{\bf 15. Institutional review board (IRB) approvals or equivalent for research with human subjects}

{\bf Question:} Does the paper describe potential risks incurred by study participants, whether such risks were disclosed to the subjects, and whether Institutional Review Board (IRB) approvals (or an equivalent approval/review based on the requirements of your country or institution) were obtained?

{\bf Answer:} \answerNA{} % Replace by \answerYes{}, \answerNo{}, or \answerNA{}.

{\bf Justification:} This paper does not involve crowdsourcing or research with any human subjects.{}

{\bf Guidelines:}
    \begin{packed_itemize}
        \item The answer NA means that the paper does not involve crowdsourcing nor research with human subjects.
        \item Depending on the country in which research is conducted, IRB approval (or equivalent) may be required for any human subjects research. If you obtained IRB approval, you should clearly state this in the paper. 
        \item We recognize that the procedures for this may vary significantly between institutions and locations, and we expect authors to adhere to the NeurIPS Code of Ethics and the guidelines for their institution. 
        \item For initial submissions, do not include any information that would break anonymity (if applicable), such as the institution conducting the review.
    \end{packed_itemize}

{\bf 16. Declaration of LLM usage}

{\bf Question:} Does the paper describe the usage of LLMs if it is an important, original, or non-standard component of the core methods in this research? Note that if the LLM is used only for writing, editing, or formatting purposes and does not impact the core methodology, scientific rigorousness, or originality of the research, declaration is not required.
%this research? 

{\bf Answer:} \answerYes{} % Replace by \answerYes{}, \answerNo{}, or \answerNA{}.

{\bf Justification:} \sreagent{} is fundamentally an LLM-based multi-agent system for autonomous Site Reliability Engineering (SRE) of cloud services. 
% The agent leverages LLMs as the core of agents (e.g., diagnosis, mitigation), each responsible for critical SRE tasks such as failure detection, root cause analysis, and recovery. 
The LLMs are central to the system's planning, reasoning, and interaction with the environment 
    through natural-language interfaces, making their usage both original and essential to the system's methodology.{}

{\bf Guidelines:}
    \begin{packed_itemize}
        \item The answer NA means that the core method development in this research does not involve LLMs as any important, original, or non-standard components.
        \item Please refer to our LLM policy (\url{https://neurips.cc/Conferences/2025/LLM}) for what should or should not be described.
    \end{packed_itemize}

\newpage
\appendix
\addcontentsline{toc}{section}{Appendix} 
\doparttoc                 
\faketableofcontents 

\part{Appendix} 
In the appendix, we provide additional analysis, 
    detailed results, and more discussions.
\vspace*{1.5cm}
\parttoc 

\clearpage
\section{Confinement Rules}
To ensure safety and predictability, each \sreagent{} agent operates 
    under sandboxed confinement rules. 
As detailed in \Cref{tab:confinement}, 
    detection and diagnosis agents are restricted 
    to a Read-Only role and are only permitted 
    to \textit{observe} the system using commands such as 
    \texttt{\small TelemetryCollection} tools. 
These agents are explicitly prohibited 
    from executing any operation 
    that alters the state of the system.

To perform mitigation, \sreagent{} will need to use the Writer role,
    as changing the system state is essential for mitigation.
    For state-changing tasks, \sreagent{} performs 
    command-level validation to verify that generated actions are 
    syntactically correct and safe to run in production. 
    Risky operations such as deleting namespaces are prohibited.
    
Certain \texttt{\small kubectl} commands, such as \texttt{\small kubectl exec -it} 
    and \texttt{\small kubectl edit}, launch interactive shells or editors. 
These commands are prevented because \sreagent{} 
    cannot participate in interactive sessions, 
    which would cause execution to hang. 
Additionally, granting arbitrary shell access significantly increases 
    the potential for unintended and unsafe behaviors. 
Therefore, we explicitly disallow all interactive commands.

\begin{table}[h!]
    \centering
    \scriptsize
    \definecolor{allowedcolor}{RGB}{230,255,230}
    \definecolor{blockedcolor}{RGB}{255,230,230}
    \caption{Agent confinement rules in \sreagent{}. \textcolor{allowedcolor}{\rule{0.3cm}{0.3cm}} shows 
        the allowed sandboxing rules according to the agent's role, 
        while \textcolor{blockedcolor}{\rule{0.3cm}{0.3cm}} shows the blocked actions enforced by \sreagent{}.}
    \label{tab:confinement}
    
    \begin{tabular}{@{}p{2.5cm} p{2.5cm} p{2.5cm} p{4.5cm}@{}}
    \toprule
    \textbf{Category} & \textbf{Subcategory} & \textbf{Confinement} & \textbf{Example(s)} \\
    \midrule
    
    \multirow{5}{*}{\textbf{Agent Role-Based}} & 
    \cellcolor{allowedcolor} & 
    \cellcolor{allowedcolor}Non-mutating & \cellcolor{allowedcolor}\texttt{kubectl get}, \texttt{kubectl describe} \\
    
    & \cellcolor{allowedcolor} & \cellcolor{allowedcolor}commands only & \cellcolor{allowedcolor}\texttt{kubectl logs}, \texttt{ls}, \texttt{cat} \\
    
    & \multirow{-3}{*}{\cellcolor{allowedcolor} \textcolor{black} Read-Only Agents} & \cellcolor{allowedcolor} & \cellcolor{allowedcolor}\texttt{TelemetryCollection} \\
    \cline{2-4}
    
    & \cellcolor{allowedcolor} & 
    \cellcolor{allowedcolor}Sequential execution & \cellcolor{allowedcolor}Only serialized state modifications \\
    
    & \cellcolor{allowedcolor}\multirow{-2}{*}{Writer Agents}  & \cellcolor{allowedcolor}only & \cellcolor{allowedcolor}are allowed \\
    
    \midrule
    \multicolumn{4}{@{}l}{\textbf{Command-Level Confinement (Blocked for all agents):}} \\
    \midrule
    
    \multirow{5}{*}{\textbf{Kubectl Subcommands}} & 
    \cellcolor{blockedcolor}Destructive Operations & \cellcolor{blockedcolor}Namespace deletion & \cellcolor{blockedcolor}\texttt{kubectl delete namespace my-ns} \\
    \cline{2-4}
    
    & \cellcolor{blockedcolor} & \cellcolor{blockedcolor}\texttt{debug}, & \cellcolor{blockedcolor}\texttt{kubectl debug pod/foo} \\
    
    & \cellcolor{blockedcolor}\multirow{-2}{*}{Interactive Edit} & \cellcolor{blockedcolor}\texttt{edit} & \cellcolor{blockedcolor}\texttt{kubectl edit deployment/bar} \\
    \cline{2-4}
    
    & \cellcolor{blockedcolor}Input from stdin & \cellcolor{blockedcolor}\texttt{-f -} pattern & \cellcolor{blockedcolor}\texttt{kubectl apply -f -} \\
    
    \midrule
    \multirow{2}{*}{\textbf{Kubectl Flags}} & 
    \cellcolor{blockedcolor} & \cellcolor{blockedcolor}\texttt{--stdin}, \cellcolor{blockedcolor}\texttt{--tty}, & \cellcolor{blockedcolor}\texttt{kubectl attach --tty pod/foo} \\
    
    & \cellcolor{blockedcolor}\multirow{-2}{*}{Interactive Terminal}  & \cellcolor{blockedcolor}\texttt{exec -it} & \cellcolor{blockedcolor}\texttt{kubectl exec -it pod/bash} \\
    
    \midrule
    \multirow{7}{*}{\textbf{Shell Syntax}} & 
    \cellcolor{blockedcolor}Pipelines & \cellcolor{blockedcolor}Pipe operator & \cellcolor{blockedcolor}\texttt{kubectl get pods | grep Running} \\
    \cline{2-4}
    
    & \cellcolor{blockedcolor}Compound Commands & \cellcolor{blockedcolor}\texttt{\&\&}, \cellcolor{blockedcolor}\texttt{||}, \cellcolor{blockedcolor}\texttt{;} & \cellcolor{blockedcolor}\texttt{kubectl get pods \&\& echo success} \\
    \cline{2-4}
    
    & \cellcolor{blockedcolor} & 
    \cellcolor{blockedcolor} & \cellcolor{blockedcolor}\texttt{kubectl delete pod \$(kubectl get} \\
    
    & \cellcolor{blockedcolor}\multirow{-2}{*}{Command Substitution} & \cellcolor{blockedcolor}\multirow{-2}{*}{\texttt{\$(...)}, backticks} & \cellcolor{blockedcolor}\texttt{pods -o name)} \\
    \cline{2-4}
    
    & \cellcolor{blockedcolor} & 
    \cellcolor{blockedcolor}\texttt{if}, \cellcolor{blockedcolor}\texttt{for}, \cellcolor{blockedcolor}\texttt{while}, & \cellcolor{blockedcolor}\texttt{if [ ... ]; then ...; fi} \\
    
    & \cellcolor{blockedcolor}\multirow{-2}{*}{Flow Control} & \cellcolor{blockedcolor}\texttt{until}, \cellcolor{blockedcolor}\texttt{case} & \cellcolor{blockedcolor}\texttt{for pod in ...; do ...; done} \\
    \cline{2-4}
    
    & \cellcolor{blockedcolor}Shell Functions & \cellcolor{blockedcolor}Function definitions & \cellcolor{blockedcolor}\texttt{myfunc() \{ ... \}} \\
    
    \bottomrule
    \end{tabular}
    \end{table}

\definecolor{PastelGreen}{rgb}{0.6, 0.8, 0.6} 
\definecolor{PastelBlue}{rgb}{0.6, 0.7, 0.8} 

\begin{mdframed}[
    linecolor=black,linewidth=1pt,topline=true,roundcorner=2pt,
    frametitlerule=true,nobreak=true,
    apptotikzsetting={\tikzset{mdfframetitlebackground/.append style={%
    shade,left color=PastelGreen, right color=PastelBlue}}},
    frametitle={\textbf{Confinement Examples}}
]
    \textbf{Example 1} (Input from stdin):

    \begin{tabular}{p{7em}|>{}p{29em}}
        \hline
        1. \sreagent{}: & \footnotesize \texttt{kubectl apply -f -} \\
        2. Confinement: & \footnotesize \textcolor{purple}{\texttt{"Stdin redirection is not allowed."}} \\
        \hline
    \end{tabular}

    \vspace{0.5em} 

    \textbf{Example 2} (Interactive Terminal):

    \begin{tabular}{p{7em}|>{}p{29em}}
        \hline
        1. AOL-agent: & \footnotesize \texttt{kubectl exec -it mongodb-rate-bfbcf4587-md6xl -n test-hotel-reservation -{}- mongo} \\
        2. AOL-agent: & \textcolor{red}{Hang there until timeout.} \\
        \hline \hline
        1. \sreagent{}: & \footnotesize \texttt{kubectl exec -it mongodb-rate-bfbcf4587-j7lsf -n test-hotel-reservation -{}- mongo} \\
        2. Confinement: & \footnotesize \textcolor{purple}{\texttt{"Interactive flag detected: -it. Such commands are not supported."}} \\
        \hline
    \end{tabular}
\end{mdframed}

To further enforce safety, \sreagent{} uses static linting to validate 
    command syntax before execution. This validation includes rejecting 
    shell pipes, compound commands (e.g., using \texttt{\small \&\&}, \texttt{\small ||}, or semicolons), 
    and other complex shell constructs. 
    These are disallowed for two reasons: 
    (1) there are exponentially many arbitrary shell expressions that \sreagent{}
        may generate; comprehensively verifying the safety of all possible
        shell expressions is not realistic. 
        On the contrary, individual \texttt{\small kubectl} invocations 
        can be easily checked. 
    (2) we require the agent to operate 
    incrementally, issuing and validating one command at a time, to work 
    effectively with our safety infrastructure and confinement layers.

When supported, \sreagent{} employs Kubernetes' dry-run mechanism 
    to produce the effect of a command without persisting changes. 
In this mode, the Kubernetes API server processes the request and 
    returns the result without modifying the system. 
This simulated output is presented to the agent, 
    which can then decide whether to proceed, 
    adjust the command, or cancel the operation entirely. 
This approach allows the mitigation agent 
    to preview the impact of state-changing actions 
    and apply corrections before execution.

The examples above show the effectiveness of 
    command-level confinement in the 
    mitigation agent of \sreagent{}, 
    particularly for preventing execution failures 
    or system hangs due to unsafe operations.
In Example 1, the agent attempts 
    to execute the command \code{kubectl} \texttt{\small apply} \code{-f} \code{-},
    which will make the agent hang due to missing input via stdin.
However, it is prevented by \sreagent{} with feedback.
In Example 2, the agent issues an interactive command 
    using the \texttt{-it} flag, 
    attempting to launch a terminal session within a running pod. 
Without confinement, this results in the execution hanging until
    a timeout occurs.
\sreagent{}'s confinement mechanism blocks such interactive terminal invocations, 
    returning a clear message to the agent:
    \texttt{\small "Interactive} \texttt{\small flag} \texttt{\small detected:} \texttt{\small -it.} \texttt{\small Such} \texttt{\small commands} \texttt{\small are} \texttt{\small not} \texttt{\small supported."} 

In our evaluation, 10.5\% of commands 
    executed by \sreagent{} (GPT-4o) 
    and 9.7\% by \sreagent{} (LLaMA3) would 
    result in system hangs, in the absence of our confinement mechanisms.
    
\section{Bounded Risk Window}
\label{para:increasing_steps}
\begin{figure}[htbp]
    \centering
        \includegraphics[width=0.75\textwidth]{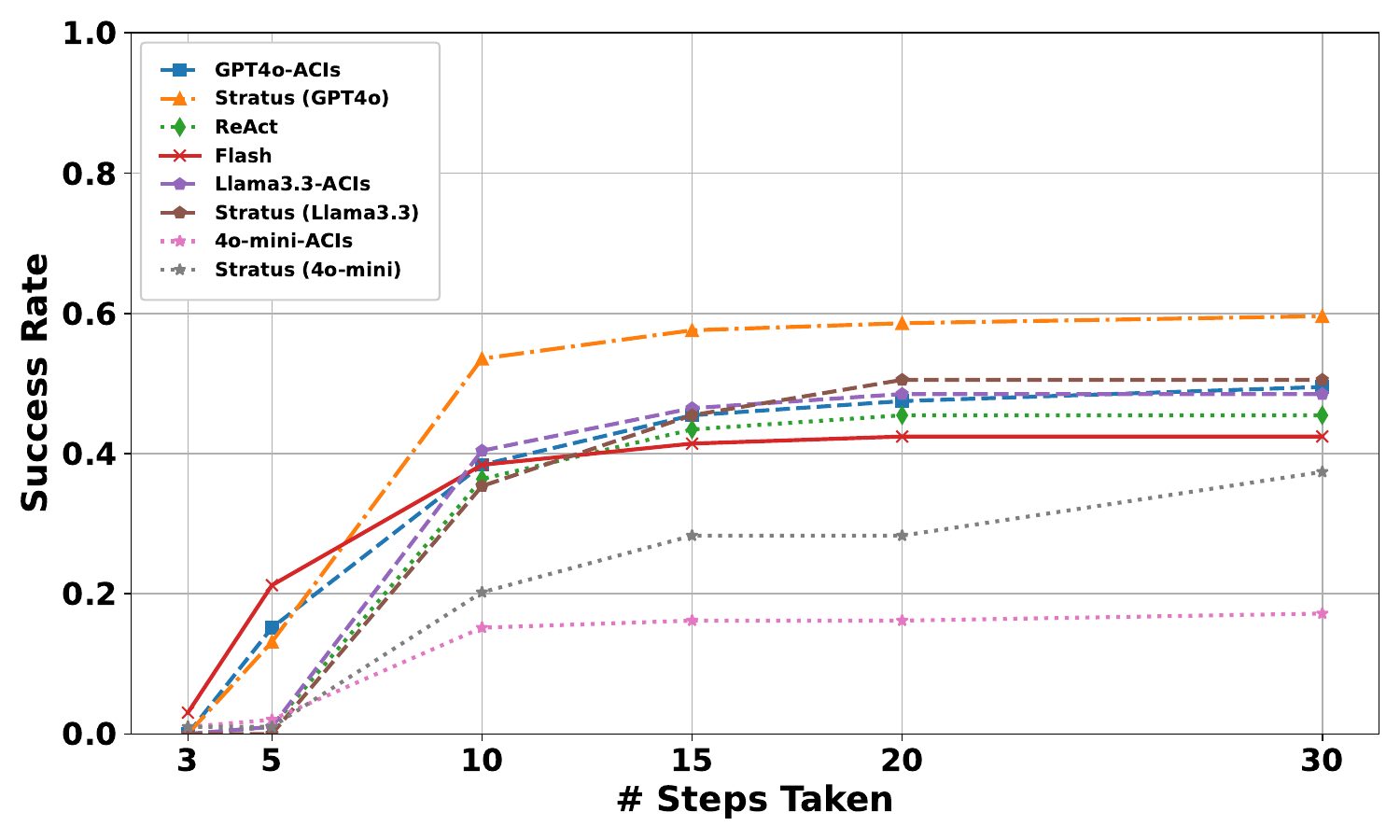}
        \caption{Impact of step limit on success rate of different agents in AIOpsLab.}
        \vspace{-10pt}
        \label{fig:k_steps}
\end{figure}

A key safety mechanism in \sreagent{} 
    is the \textit{bounded risk window},
    which specifies the number of actions
    that the mitigation agent ($\alpha_M$) 
    can execute in a single transaction.
In this section, we empirically investigate 
    how the total step limit impacts \sreagent{}'s 
    success rate on AIOpsLab problems
    (detection, localization, root-cause analysis, and mitigation).
For each problem, we run the evaluated agents 
    with total step limits setting to different values 
    (3, 5, 10, 15, 20, and 30).
Results are presented in \Cref{fig:k_steps}.

\sreagent{} (GPT-4o) improves from 0\% to 53.53\%,
    after switching the step limit from 3 to 10.
This shows that allowing \sreagent{} 
    to take more steps can significantly
    boost its effectiveness.
We also find that among all evaluated agents, 
    agent performance plateaus after 15 steps, 
    indicating that taking more steps 
    after a certain threshold 
    does not yield better performance significantly.  

The effectiveness gain can be observed 
    on \sreagent{} with other LLM backends,
    but with lower overall accuracy, relatively scaling with
    the LLM's capability.  
Based on these observations, 
    we set the default bounded risk window to $K=20$.

\section{Ablation Study on ITBench}

In \S\ref{sec:undo-and-retry}, we presented an ablation study of 
    \sreagent{} (GPT-4o) on AIOpsLab, 
    demonstrating the effectiveness of its retry and undo mechanisms. 
We now examine \sreagent{}'s behavior under ablation 
    on the ITBench benchmark,
    as shown in \Cref{table:itbench_ablation_mitigation}.
We observe that (1) disabling retry 
    substantially drops the success rate to 11.1\%. 
Note that this configuration implicitly disables the undo agent, 
    as there are no prior actions to undo.
Without retries, the agent must determine the root cause 
    and devise a mitigation plan correctly with one attempt, 
    which is significantly challenging.
(2) \sreagent{} achieves the same task-level success rate 
    (solving 9/18 problems) when the undo agent is disabled. 
As mentioned in \S\ref{sec:failure-mitigation-effectiveness}, 
    \sreagent{} exploits the observation that in
    8 out of 18 problems,
    restarting the target pods clears the incident alerts. 
Since ITBench evaluates correctness primarily 
    based on the absence of immediate alerts, 
    this allows the agent to resolve the incident 
    without trying more destructive, 
    state-changing actions, 
    where our undo mechanism will be most effective.

To better understand this, we analyzed these 8 problems in detail. 
Among all mitigation actions issued by the agent, 
    approximately 15.2\% are pod deletion operations 
    managed by Kubernetes Deployment, 
    which can automatically recreate the pods. 
Furthermore, a significant portion (44.2\%) of state-changing commands,
    e.g., changing a container image to an image with a hallucinated name,
    are easily detected and fixed by \sreagent{} in the subsequent run,
    since \sreagent{} has a reflection mechanism, 
    which can help the agent to realize what went wrong in the previous run.
The rest of the commands are a mix of direct restart or redeployment of
    the pods, which again lifts the underlying fault directly.
In this regard, the undo mechanism 
    does not improve the success rate in ITBench.
However, the absence of undo capabilities 
    can still compromise the cloud over time. 
Latent or non-critical side effects from agent actions, 
    such as misconfigurations or performance degradations 
    that do not immediately trigger alerts,
    may go undetected. 
Without TNR, there is no guarantee 
    ensuring that such side effects are identified and reverted.

\begin{table}[htbp]
    \centering
    \caption{\textbf{Ablation analysis of \sreagent{} (GPT-4o) on mitigation problems in ITBench.}}
    \footnotesize
    \setlength{\tabcolsep}{2.5pt}
    \vspace{3.5pt}
    \begin{tabular}{lccc}
            \toprule
            \textbf{Ablation} & \textbf{Succ. Rate} & \textbf{Time (s)} & \textbf{Cost (\$)} \\
            \midrule
            \sreagent{} (4o) & 50\% & 1720.8 & 6.11 \\
            $-$ No Retry & 11.1\% & 267.2 & 0.72 \\
            $-$ Na\"{i}ve retry w/o undo & 50\% & 1711.9 & 5.13 \\
            \bottomrule
        \end{tabular}
    \label{table:itbench_ablation_mitigation}
\end{table}

\section{Tool Usage}
We develop a set of tools 
    that enable agents 
    to interact with the cloud environment
    through natural language.  
We describe the implemented tools in \S\ref{sec:tool-list}, 
    and analyze their usage distribution 
    across execution steps in \S\ref{sec:tool-usage}.

\subsection{List of Tools}
\label{sec:tool-list}
In order to do complex reasoning and safe system manipulation,
    \sreagent{} has a comprehensive suite of tools,
    including observability tool and command-line tools.
We describe some of the tools here.

\paragraph{Observability Tool.} 
\sreagent{} has tools to observe the clouds,
    through collecting telemetry data (\code{TelemetryCollection}), 
    or retrieving system states.

\begin{itemize}
    \item \code{GetLogsTool}: Retrieves logs from specific component, e.g., application logs from a given pod.
    \item \code{GetTracesTool} and \code{ReadTracesTool}: Retrieve traces if the system is traced, e.g., traces from Jaeger. \code{ReadTracesTool} handles large trace files stored on disk.
    \item \code{GetBootstrapAnalysisTool} and \code{ReadBootstrapAnalysisTool}: Support trace-based bootstrapping by identifying likely fault locations. Instead of digesting the entire trace data, the agent will only read the traces with errors (e.g., \code{500} HTTP responses) as initial analysis results.
    \item \code{OracleTools}: After each iteration 
        of diagnosis or mitigation, oracles can be used to validate system health 
        (e.g., alert clearance, workload throughput, cluster health). 
        These oracles determine whether the agent should continue, retry, or terminate. 
\end{itemize}
Note that \sreagent{} disables direct metrics consumption, 
    as raw numerical data can be difficult 
    for the agent to interpret effectively. 
Instead, such information is surfaced in logs 
    in a more pre-processed form, 
    e.g., as filtered, aggregated, or human-readable summaries.
These tools abstract away raw observability data, 
    enabling the agent to reason over high-level, 
    interpretable signals,
    and to determine its termination.    
\paragraph{Command-line Tools.}
\label{sec:tool-usage}
On the action side, 
    \sreagent{} has specialized tools to construct the command 
    to change the system states and perform the failthful undo.
\begin{itemize}
    \item \code{NL2Kubectl}: Translates natural language instructions generated by the agent into Kubernetes commands. This decouples high-level reasoning from low-level execution syntax. 
    \item \code{LintingTool}: Acts as an internal tool to check the correctness of the command, e.g., the commands generated by \code{NL2Kubectl}; if the command is not correct or allowed, the tool will return a message to ask the agent to improve the command.
    \item \code{RollbackTool}: Supports the undo agent to perform failthful undo actions.
    \item \code{NotificationTool}: Whenever the agents feel confident that the task is finished, it can use the \code{NotificationTool} to notify human of task completion, or escalation. Notification can be made in different troubleshooting phases e.g., detection phase should notify human whether there is a nomaly in the system; while the agent should notify human whether the mitigation is successful or complete.
\end{itemize}

\subsection{Distribution of Tool Usage}
\begin{figure}[htbp]
        \centering
        \includegraphics[width=0.70\textwidth]{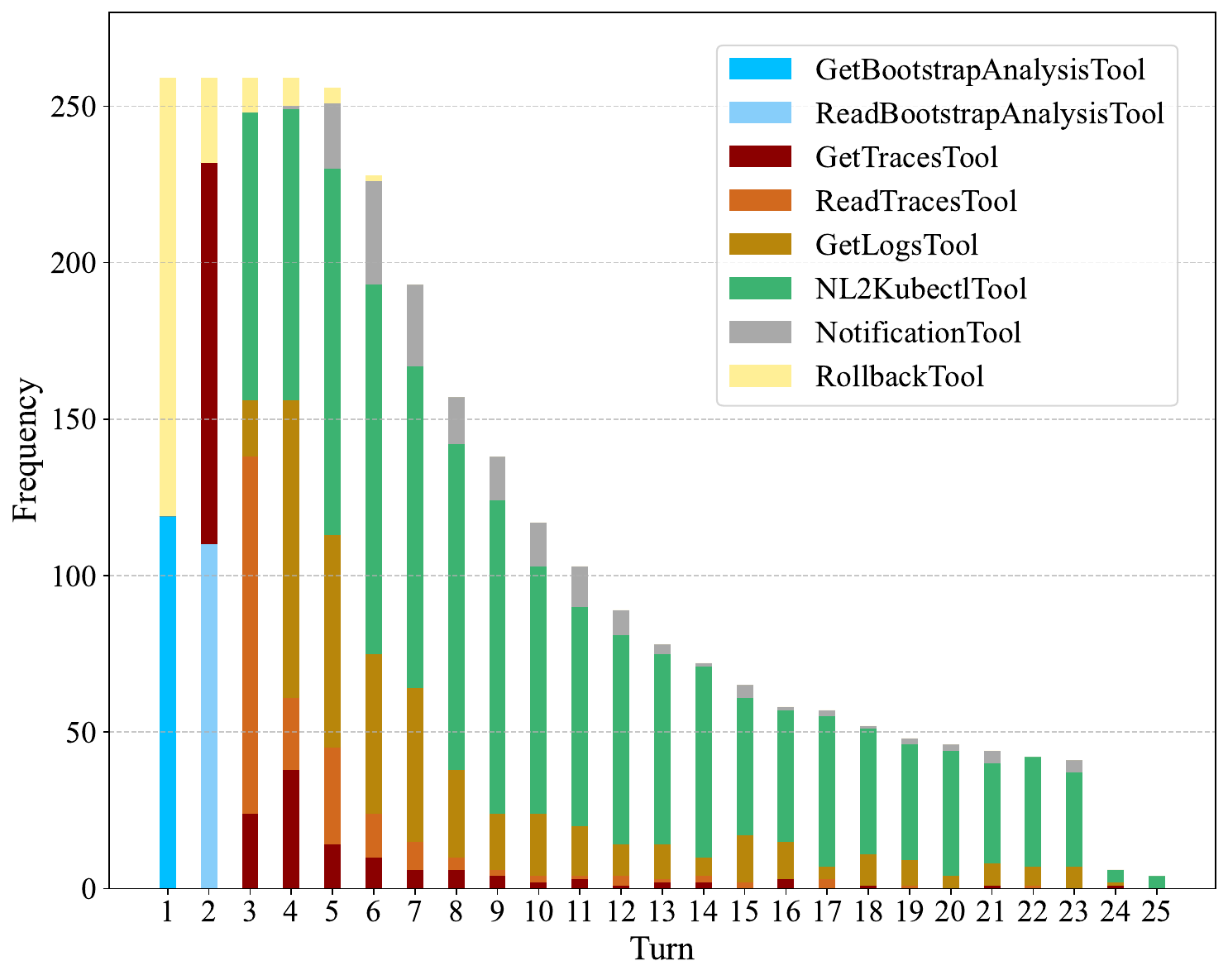}
        \caption{Distribution of tool usage by \sreagent{} (GPT-4o) across steps.
        The figure shows a histogram of tool invocations grouped by type and step index (Iteration).
        }
        \label{fig:tool_usage_chart}
\end{figure}

To analyze the tool invocation patterns in \sreagent{}, 
    we log the tool usage of the \sreagent{} (GPT-4o)
    across all problems in the AIOpsLab.
For each execution step (i.e., each iteration 
    within the retry loop), 
    we record which tools 
    that interact with the clouds 
    are invoked.
In \Cref{fig:tool_usage_chart}, we show the distribution of external tool usage (including the \code{TelemetryCollection}, \code{NL2Kubectl}, \code{NotificationTool}, and \code{RollbackTool}) by the \sreagent{} (GPT-4o) at different steps in our AIOpsLab experiments. 

\sreagent{} exhibits 
    distinct phases of tool usage across execution steps. 
In the initial stage (Iterations 1--2), 
    the \code{RollbackTool} is predominantly used by the undo agent, 
    and bootstrapping tools are used to perform initial analysis,
    which are specified in the control flow. 
In early steps (Iterations 3--8), 
    the agent primarily engages \code{TelemetryCollection} tools 
    to gather observability data from logs and traces. 
As execution progresses (Iterations 9--23), 
    \code{NL2KubectlTool} dominates, 
    indicating the agent's transition 
    from information gathering 
    using provided tools
    to more customized command generation 
    for diagnosis and mitigation.
Additionally, we observe a shift in preference 
    from trace retrieval to log retrieval over time.

\section{Oracle Usage}

In \S\ref{sec:term-oracle}, 
    we describe how \sreagent{} determines termination
    based on the oracles.
Here, we use an example (Problem ID: \code{wrong_bin_usage-mitigation-1} in AIOpslab) 
    to illustrate the usage of our oracles in \sreagent{}.
In this problem, a microservice is misconfigured 
    to use an incorrect binary during container initialization. 
As a result, while the container is able to start 
    and all pods appear healthy at the orchestration level, 
    the application has functional failures. 
Specifically, the service responds
    with a high rate of HTTP errors (non-2xx/3xx responses), 
    indicating that the application is not working as expected.

\begin{figure}[h!]
    \centering
    \includegraphics[width=0.65\linewidth]{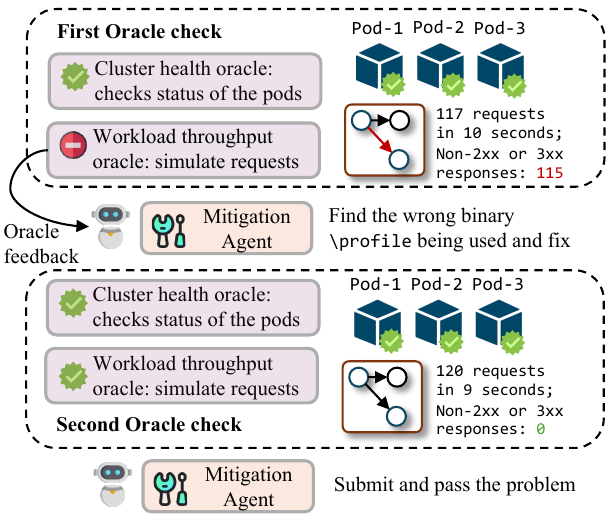}
    \caption{Oracle usage when \sreagent{} solving the \code{wrong_bin_usage-mitigation-1} task.
    In the first validation round, the workload oracle detects 115 failed responses, 
    indicating the issue is unresolved. \sreagent{} continues troubleshooting and patches the deployment with the correct binary. In the second round, both oracles confirm success.
    }
    \label{fig:oracle_example}
\end{figure}

As illustrated in \Cref{fig:oracle_example}, 
    during the initial validation phase, 
    the system health oracle confirms 
    that all pods are running normally. 
However, the workload oracle detects 
    a number of erroneous responses (115), 
    revealing that the issue has not been fully resolved. 
Based on this feedback, 
    \sreagent{} further troubleshoots the issue,
    and later correctly identifies the root cause 
    as an incorrect binary and proceeds 
    to patch the deployment with the appropriate version.
A second round of oracle validation shows that all checks pass: 
    the cluster remains healthy 
    and the workload oracle confirms the elimination of HTTP errors 
    (zero non-2xx/3xx responses). 
With successful validation, \sreagent{} notifies the benchmark, 
    marking the resolution of the issue.

\newpage
\section{Hyperparameter Sweep}
We perform a hyperparameter sweep 
    over the temperature parameter in \sreagent{} (GPT-4o)
    using all 13 mitigation tasks from AIOpsLab.
\Cref{table:temperature_mitigation} shows the results.
We can see that the temperature of 0.0
    yields the best results.
Setting temperature to 0 allows the LLM to output deterministically with a
    given input. 
Setting a non-zero temperature allows ``creativity'' in LLM's output.
However, in the SRE context, 
    this may not be desirable.
A non-zero temperature permits the LLM 
    to output non-deterministic 
    results given the same input.
For example, given a microservice application trace that clearly indicates
    the root cause of a problem, we would like \sreagent{} to confidently
    take \textit{one} correct mitigation action. 
In addition, setting overly high temperature can directly affect
    the tool's ability in generating the correct commands, which harms
    \sreagent{}'s performance critically.

\begin{table}[htbp]
    \centering
    \footnotesize
    \caption{\textbf{Effect of Temperature on Mitigation Success Rate.}}
    \label{table:temperature_mitigation}
    \begin{tabular}{ccccc}
      \toprule
      \textbf{Temperature} & \textbf{Succ.} & \textbf{Time (s)} & \textbf{Step} & \textbf{\$} \\
      \midrule
      0.0 & \cellcolor{mellowgreen!100} 69.2\% & 811.9 & 46.3 & 0.877 \\
      0.2 & \cellcolor{mellowgreen!74 } 51.3\% & 961.9 & 71.6 & 0.647 \\
      0.4 & \cellcolor{mellowgreen!67 } 46.2\% & 950.58 & 67.9 & 0.493 \\
      0.6 & \cellcolor{mellowgreen!82 } 56.4\% & 836.8 & 55.1 & 0.388 \\
      0.8 & \cellcolor{mellowgreen!48 } 33.3\% & 1209.51 & 77.2 & 0.533 \\
      \bottomrule
    \end{tabular}
\end{table}
\section{Result Details}

We present detailed evaluation results 
    for \sreagent{} (GPT-4o) in AIOpsLab. 
\Cref{table:detailed_results} shows the results of 
    detection, localization, and RCA tasks.
\Cref{table:detailed_results_mitigation} 
    shows the mitigation results.

\definecolor{allowedcolor}{RGB}{230,255,230}
\definecolor{blockedcolor}{RGB}{255,230,230}
\newcommand{\resultTrue}{\cellcolor{allowedcolor}T}
\newcommand{\resultFalse}{\cellcolor{blockedcolor}F}

\begin{table}[htbp]
    \centering
    \scriptsize
    \vspace{-10pt}
    \caption{Detailed results of Detection, Localization, and RCA for \sreagent{} (GPT-4o) in AIOpsLab. We use the Pass@3 metric to calculate localization accuracy.
    Non-existent tasks are denoted as \texttt{--}.
    }
    \setlength{\tabcolsep}{2pt}
    \label{table:detailed_results}
    \begin{tabular}{lcccccccccccc}
      \toprule
      & \multicolumn{3}{c}{\bf Detection} & & \multicolumn{4}{c}{\bf Localization} & & \multicolumn{3}{c}{\bf RCA} \\
      \cline{2-4}
      \cline{6-9}
      \cline{11-13}
      \textbf{Problem ID}  & \textbf{Time (s)} & \textbf{Pass} & \textbf{\$} & & \textbf{Time (s)} & \textbf{Pass@3} & \textbf{Succ.} & \textbf{\$} & & \textbf{Time (s)} & \textbf{Succ.} & \textbf{\$} \\
      \midrule
\textbf{\tiny k8s\_target\_port-misconfig-1                           } & 39.04 & \resultTrue{}  & 0.100 & & 29.90 & \resultTrue{}  & 100  & 0.093 & & 49.33 & 0\% & 0.073 \\
\textbf{\tiny k8s\_target\_port-misconfig-2                           } & 49.46 & \resultTrue{}  & 0.176 & & 27.97 & \resultTrue{}  & 100  & 0.088 & & 28.97 & 50\%  & 0.089 \\
\textbf{\tiny k8s\_target\_port-misconfig-3                           } & 42.46 & \resultTrue{}  & 0.141 & & 41.19 & \resultTrue{}  & 100  & 0.147 & & 41.00 & 50\%  & 0.075 \\
\textbf{\tiny auth\_miss\_mongodb                                     } & 42.26 & \resultTrue{}  & 0.117 & & 28.70 & \resultTrue{}  & 50   & 0.115 & & 43.30 & 0\% & 0.083 \\
\textbf{\tiny revoke\_auth\_mongodb-1                                 } & 42.98 & \resultTrue{}  & 0.066 & & 58.26 & \resultTrue{}  & 100  & 0.050 & & 37.05 & 50\% & 0.042 \\
\textbf{\tiny revoke\_auth\_mongodb-2                                 } & 45.04 & \resultTrue{}  & 0.089 & & 33.27 & \resultTrue{}  & 100  & 0.041 & & 45.45 & 50\% & 0.056 \\
\textbf{\tiny user\_unregistered\_mongodb-1                           } & 35.63 & \resultTrue{}  & 0.043 & & 34.47 & \resultTrue{}  & 100  & 0.041 & & 34.10 & 50\% & 0.042 \\
\textbf{\tiny user\_unregistered\_mongodb-2                           } & 38.21 & \resultTrue{}  & 0.044 & & 34.26 & \resultTrue{}  & 100  & 0.042 & & 35.72 & 50\% & 0.042 \\
\textbf{\tiny misconfig\_app\_hotel\_res                              } & 45.74 & \resultTrue{}  & 0.095 & & 36.91 & \resultTrue{}  & 100  & 0.029 & & 45.97 & 100\%  & 0.058 \\
\textbf{\tiny scale\_pod\_zero\_social\_net                           } & 72.04 & \resultTrue{}  & 0.113 & & 35.15 & \resultTrue{}  & 100  & 0.110 & & 42.39 & 0\% & 0.076 \\
\textbf{\tiny assign\_to\_non\_existent\_node\_social\_net            } & 28.62 & \resultTrue{}  & 0.045 & & 28.91 & \resultTrue{}  & 100  & 0.045 & & 28.54 & 0\% & 0.046 \\
\textbf{\tiny container\_kill                                         } & 43.77 & \resultTrue{}  & 0.042 & & 39.72 & \resultTrue{}  & 100  & 0.063 & & --    & --                     & --    \\
\textbf{\tiny pod\_failure\_hotel\_res                                } & 68.94 & \resultTrue{}  & 0.050 & & 42.97 & \resultTrue{}  & 100  & 0.064 & & --    & --                     & --    \\
\textbf{\tiny pod\_kill\_hotel\_res                                   } & 55.75 & \resultTrue{}  & 0.072 & & 275.83& \resultFalse{} & 0    & 0.120 & & --    & --                     & --    \\
\textbf{\tiny network\_loss\_hotel\_res                               } & 43.17 & \resultTrue{}  & 0.061 & & 58.97 & \resultFalse{} & 0    & 0.070 & & --    & --                     & --    \\
\textbf{\tiny network\_delay\_hotel\_res                              } & 38.83 & \resultTrue{}  & 0.082 & & 35.87 & \resultFalse{} & 0    & 0.081 & & --    & --                     & --    \\
\textbf{\tiny noop\_detection\_hotel\_reservation                     } & 38.62 & \resultFalse{} & 0.101 & & --    & --                       & --   & --    & & --    & --                     & --    \\
\textbf{\tiny noop\_detection\_social\_network                        } & 38.10 & \resultFalse{} & 0.052 & & --    & --                       & --   & --    & & --    & --                     & --    \\
\textbf{\tiny astronomy\_shop\_ad\_service\_failure                   } & 87.84 & \resultTrue{}  & 0.094 & & 67.51 & \resultFalse{} & 0    & 0.067 & & --    & --                     & --    \\
\textbf{\tiny astronomy\_shop\_ad\_service\_high\_cpu                 } & 40.44 & \resultTrue{}  & 0.060 & & 67.16 & \resultTrue{}  & 50   & 0.071 & & --    & --                     & --    \\
\textbf{\tiny astronomy\_shop\_ad\_service\_manual\_gc                } & 42.17 & \resultTrue{}  & 0.086 & & 64.73 & \resultFalse{} & 0    & 0.055 & & --    & --                     & --    \\
\textbf{\tiny astronomy\_shop\_cart\_service\_failure                 } & 38.37 & \resultTrue{}  & 0.092 & & 60.96 & \resultTrue{}  & 100  & 0.111 & & --    & --                     & --    \\
\textbf{\tiny astronomy\_shop\_image\_slow\_load                      } & 51.59 & \resultTrue{}  & 0.072 & & 76.70 & \resultTrue{}  & 33.33& 0.209 & & --    & --                     & --    \\
\textbf{\tiny astronomy\_shop\_kafka\_queue\_problems                 } & 66.00 & \resultTrue{}  & 0.126 & & 65.48 & \resultFalse{} & 0    & 0.059 & & --    & --                     & --    \\
\textbf{\tiny astronomy\_shop\_payment\_service\_failure              } & 65.14 & \resultTrue{}  & 0.054 & & 67.03 & \resultFalse{} & 0    & 0.085 & & --    & --                     & --    \\
\textbf{\tiny astronomy\_shop\_payment\_service\_unreachable          } & 52.18 & \resultTrue{}  & 0.065 & & 66.44 & \resultFalse{} & 0    & 0.045 & & --    & --                     & --    \\
\textbf{\tiny astronomy\_shop\_product\_catalog\_service\_failure     } & 36.90 & \resultTrue{}  & 0.084 & & 62.56 & \resultFalse{} & 0    & 0.046 & & --    & --                     & --    \\
\textbf{\tiny astronomy\_shop\_recommendation\_service\_cache\_failure} & 78.68 & \resultTrue{}  & 0.101 & & 54.93 & \resultFalse{} & 0    & 0.090 & & --    & --                     & --    \\
\textbf{\tiny astronomy\_shop\_loadgenerator\_flood\_homepage         } & 29.59 & \resultTrue{}  & 1.183 & & 35.45 & \resultFalse{} & 0    & 1.345 & & --    & --                     & --    \\
\textbf{\tiny redeploy\_without\_PV                                   } & 40.72 & \resultTrue{}  & 0.044 & & --    & --                       & --   & --    & & 42.74 & 0\% & 0.070 \\
\textbf{\tiny wrong\_bin\_usage                                       } & 40.32 & \resultTrue{}  & 0.170 & & 298.09& \resultFalse{} & 0    & 0.149 & & 40.86 & 50\% & 0.137 \\
\textbf{\tiny noop\_detection\_astronomy\_shop                        } & 68.54 & \resultFalse{} & 0.056 & & --    & --                       & --   & --    & & --    & --                     & --    \\
      \bottomrule
\textbf{Average} & 48.35 & 29/32 & 0.118 & & 65.34 & 16/28 & 51.19 & 0.126 & & 39.65 & 34.62\% & 0.068 \\
      \bottomrule
    \end{tabular}
    \vspace{-10pt}
\end{table}

\begin{table}[htbp]
    \centering
    \footnotesize
    \caption{Detailed results of Mitigation tasks for \sreagent{} (GPT-4o) in AIOpsLab.}
    \label{table:detailed_results_mitigation}
    \begin{tabular}{lcccc}
      \toprule
      \textbf{Problem ID}  & \textbf{Time (s)} & \textbf{Step} & \textbf{Succ.} & \textbf{\$} \\
      \midrule
{\footnotesize k8s\_target\_port-misconfig-mitigation-1                  } & 124.59   & 13  & \resultTrue{}  & 0.116 \\
{\footnotesize k8s\_target\_port-misconfig-mitigation-2                  } & 210.08   & 23  & \resultTrue{}  & 0.338 \\
{\footnotesize k8s\_target\_port-misconfig-mitigation-3                  } & 148.49   & 11  & \resultTrue{}  & 0.084 \\
{\footnotesize assign\_to\_non\_existent\_node\_social\_net-mitigation-1 } & 115.42   & 12  & \resultTrue{}  & 0.051 \\
{\footnotesize scale\_pod\_zero\_social\_net-mitigation-1                } & 139.03   & 88  & \resultTrue{}  & 0.232 \\
{\footnotesize user\_unregistered\_mongodb-mitigation-1                  } & 2,007.58 & 123 & \resultTrue{}  & 0.330 \\
{\footnotesize user\_unregistered\_mongodb-mitigation-2                  } & 2,090.04 & 46  & \resultFalse{} & 0.955 \\
{\footnotesize revoke\_auth\_mongodb-mitigation-1                        } & 1,405.48 & 121 & \resultFalse{} & 0.436 \\
{\footnotesize revoke\_auth\_mongodb-mitigation-2                        } & 1,321.47 & 108 & \resultFalse{} & 0.644 \\
{\footnotesize auth\_miss\_mongodb-mitigation-1                          } & 136.60   & 20  & \resultTrue{}  & 0.148 \\
{\footnotesize misconfig\_app\_hotel\_res-mitigation-1                   } & 1,367.99 & 10  & \resultFalse{} & 0.805 \\
{\footnotesize redeploy\_without\_PV-mitigation-1                        } & 839.00   & 152 & \resultTrue{}  & 6.704 \\
{\footnotesize wrong\_bin\_usage-mitigation-1                            } & 649.44   & 51  & \resultTrue{}  & 0.562 \\
\bottomrule
\textbf{Average}                                                       & 811.94  & 59.85 & 9/13 & 0.877   \\
\bottomrule
    \end{tabular}
\end{table}

\section{Heuristic Evaluation of \sreagent's Multi-agent Design}

\sreagent{} adopts a multi-agent design.
Recent work~\cite{cemri2025multiagentllmsystemsfail} 
    categorizes common failure modes of multi-agent systems,
    including
    \emph{Specification Issues}, \emph{Inter-Agent Misalignment}, and \emph{Task Verification}.
We present a heuristic evaluation of the \sreagent{} design and 
    how the design addresses common failure modes.
\Cref{tbl:failure} provides a summary.

\begin{table}[htbp]
    \centering
    \caption{Failure modes from \cite{cemri2025multiagentllmsystemsfail} and \sreagent{} techniques that mitigate them. Abbreviations: DCF=\textsl{Deterministic~Control~Flow}, CRB=\textsl{Confinement (role-based)}, CCL=\textsl{Confinement (command-level)}, ORC=\textsl{Oracles}, STS=\textsl{Specialized~Toolset}, UDO=\textsl{Undo}, LTR=\textsl{Linter}, BTS=\textsl{Bootstrapping}, ATD=\textsl{Agent~Thought~Dropout}, RWR=\textsl{Reflection~with~Retries}. Each row shows how the technique conceptually addresses the failure.}
    \label{tbl:failure}
    {
    \scriptsize
    \begin{tabularx}{\textwidth}{p{3.5cm}|cX}
    \toprule
    \textbf{Failure Mode} & \textbf{\sreagent{} Technique(s)} & \textbf{Description and Examples} \\
    \midrule
    \multicolumn{3}{l}{\textbf{Specification Issues (System Design)}} \\
    \hline
    \multirow{2}{*}{Disobey Task Specification} & \multirow{2}{*}{CRB, CCL}  & Role-based confinement assigns specific tasks for different troubleshooting stages; command-level confinement will validate each action;  \\
    \hline
    \multirow{2}{*}{Disobey Role Specification} & \multirow{2}{*}{CRB, STS} & Each agent has clear roles (diagnosis, mitigation, etc) and is given tools corresponding to the agent type. This allows for agent-specific confinement where each agent is only given the necessary tools to complete the task it's been assigned. \\
    \hline
    \multirow{2}{*}{Step Repetition} & \multirow{2}{*}{ATD, BTS} & Agent thought dropout prevents agents from being trapped in previous misleading thoughts; bootstrapping gives focused start. \\
    \hline
    Loss of Conversation History & RWR & Persistent agent thoughts retain context across retries. \\
    \hline
    Unaware of Termination Conditions & ORC & Oracles identify task completion and provide extra context. \\
    \midrule
    \multicolumn{3}{l}{\textbf{Inter-Agent Misalignment (Agent Coordination)}} \\
    \hline
    \multirow{2}{*}{Conversation Reset} & \multirow{2}{*}{DCF, UDO} & Dataflow direction between agents is the same as control flow; undo restores state on incorrect or destructive actions. \\
    \hline
    Fail to Ask for Clarification & RWR & A new retry round can be initiated if the agent fails the task. \\
    \hline
    \multirow{3}{*}{Task Derailment} & \multirow{3}{*}{BTS, ATD, UDO} & Bootstrapping orients initial step (i.e., where the agent starts); contaminated thought can be dropped; undo can be used to restore the state. \\
    \hline
    \multirow{2}{*}{Information Withholding} & \multirow{2}{*}{DCF} & Agents will be forced to give a final answer (a part of CrewAI's design) to determine its success or to notify of failure. \\
    \hline
    Ignored Other Agent's Input & DCF, RWR & Agents consume its predecessor's output (including output of the previous round in retrying) in the dataflow until the task is completed. \\
    \hline
    Reasoning-Action Mismatch & -- & Probably related more to the model's own capabilities. \\
    \midrule
    \multicolumn{3}{l}{\textbf{Task Verification (Quality Control)}} \\
    \hline
    Premature Termination & ORC, DCF & Task ends only after all agents finish and oracles confirm completion. \\
    \hline
    No or Incomplete Verification & ORC, CRB, CCL & Task completion and single action are both verified. \\
    \hline
    Incorrect Verification & ORC & Multiple oracles provide validation to reduce false positives. \\
    \bottomrule
    \end{tabularx}
    }
    \end{table}
    
{\bf Specification Issues.}
A common failure pattern arises from incomplete 
    or ambiguous role/task definitions, 
    or insufficient system design encoding in agent prompts~\cite{cemri2025multiagentllmsystemsfail}.
\sreagent{} avoids this failure pattern in multiple aspects (see Table~\ref{tbl:failure}).
Role-based confinement (CRB) 
    is used to confine the agent's behavior 
    according to the role and task specifications; 
    command-level confinement (CCL) will validate each action.
Also, \sreagent{} has specialized toolset (STS) for different agents,
    to retrieve the relevant information from the system,
    which can facilitate their task completion.
{\bf Inter-Agent Misalignment.}
The Inter-Agent Misalignment category 
    includes coordination failures among agents, 
    such as one agent's incorrect output misleading downstream agents~\cite{cemri2025multiagentllmsystemsfail}. 
Take our context as the example. 
If the detection agent identifies the wrong error location, 
    the mitigation agent may inadvertently repair the wrong service.
\sreagent{} addresses such issues 
    by employing deterministic control flow (DCF), 
    which coordinates the agents to work together
    and communicate clearly with each other.
\sreagent{} also uses reflection with retries (RWR), 
    which enables the system to reassess and relaunch a task 
    (within the bounded risk window $K$).
In addition, the undo capability (UDO) allows the system 
    to revert to a safe state after erroneous executions. 
Furthermore, agent thought dropout (ATD) prevents 
    potentially corrupted 
    or misleading context propagated from previous agents.
{\bf Task Verification.}
Inadequate verification that fails to detect or correct errors
    or premature termination of tasks
    can make multi-agent systems fail~\cite{cemri2025multiagentllmsystemsfail}.
\sreagent{} deploys three oracles (ORC) including 
    alert oracle, 
    user request oracle, 
    and system health oracle, 
    which periodically assess system states 
    and validate whether termination conditions have been met,
    to ensure the agent is terminating appropriately.
Complementing these, \sreagent{} has confinement (CRB, CCL, LTR), e.g., 
    command dry-run and syntax check,
    to ensure that commands are semantically and syntactically 
    checked prior to execution.

\section{An Example of \sreagent{}'s Trajectory}
We show the trajectory of \sreagent{} 
    when solving a mitigation task
    (Problem ID: \code{redeploy_without_PV-mitigation-1}) in AIOpsLab.
In this problem, 
    the storageclass of the persistent volumes (PVs)
    is mistakenly removed during redeployment. 
As a result, the persistent volumes claims (PVCs) 
    will not be able to find the corresponding storageclass
    in the new deployment,
    and most of the pods will be in an erroneous state.

We show the key actions taken by \sreagent{} in ``Gold Action.'' Statements marked as ``Note'' 
    are not part of the trajectory but are added for clarity.

\begin{frameworklog}{Initialization}{1}
[21:34:50] All pods in namespace 'test-hotel-reservation' are ready.                                                               
== Fault Injection ==
[21:34:51] Waiting for namespace 'test-hotel-reservation' to be deleted...                                                           
[21:35:18] Namespace 'test-hotel-reservation' has been deleted.
Namespace 'test-hotel-reservation' deleted successfully.
Deleting namespace test-hotel-reservation without deleting the PV.
Redeploying test-hotel-reservation.
Namespace test-hotel-reservation not found. Creating namespace.
Namespace test-hotel-reservation created successfully: namespace/test-hotel-reservation created
\end{frameworklog}

\begin{agenttask}{Bootstrapping}{Diagnosis}
Analyze the traces to localize which part of the system, e.g., a service, a component, a pod, a node, etc., has anomalies. Always use the get_traces tool first; then use a read_traces tool to help you with the localization. Note that it may not be helpful to use the tools too many times. You are working on the namespace 'test-hotel-reservation', please use this namespace when using the tools.
\end{agenttask}

\begin{agentthought}{Diagnosis}
To analyze the traces and localize anomalies in the 'test-hotel-reservation' namespace, I will first use the get\_traces tool to fetch the traces for a specified duration. Then, I will use the read\_traces tool to read and analyze the traces.

\vspace{1em} \hrule \vspace{1em}

Call get\_traces with \{"namespace": "test-hotel-reservation", "duration": 30\}
\end{agentthought}

\begin{agenttool}{get\_traces}{Diagnosis}
Traces data exported to: /eval/xagent/trace_output/traces_1747186602.csv
Please take the next action
\end{agenttool}

\begin{agentthought}{Diagnosis}
Call read\_traces with \{"file\_path": "/eval/xagent/trace\_output/traces\_1747186602.csv"\}
\end{agentthought}

\begin{agenttool}{read\_traces}{Diagnosis}
The following list of services are the last service the request reached before returning an error, these services may not be the exact place of the root cause, so you may need to dig deeper. These services correspond to a list of operations happened there, which is the service relying on and could be where the root cause comes from. Here is the list:
1: {service: "search", operation: "geo"}
2: {service: "frontend", operation: "recommendation"}
\end{agenttool}

\begin{agentanswer}{Bootstrapping}{Diagnosis}
- List of entities with anomalies:
  1. Service: "search", Operation: "geo"
  2. Service: "frontend", Operation: "recommendation"

- Possible root cause entities for further analysis:
  - Service: "search", Operation: "geo"
  - Service: "frontend", Operation: "recommendation"
\end{agentanswer}

\begin{agenttask}{Mitigation}{Mitigation}
The problem description is '''You are an expert DevOps engineer assigned to mitigate anomalies in a deployed service...
\end{agenttask}

\begin{agentthought}{Mitigation}
To begin the mitigation process, I need to gather more information about the services with anomalies. The anomalies are present in the "search" service with the "geo" operation and the "frontend" service with the "recommendation" operation. I will start by fetching the logs for these services to understand the issues better.

First, I will fetch the logs for the "search" service in the "test-hotel-reservation" namespace.

\vspace{1em} \hrule \vspace{1em}

Call get\_logs with \{"namespace": "test-hotel-reservation", "service": "search"\}
\end{agentthought}

\begin{agenttool}{get\_logs}{Mitigation}
{"level":"info","time":"2025-05-14T01:36:03Z","message":"TLS disabled."}
{"level":"info","time":"2025-05-14T01:36:03Z","message":"Set global log level: info"}
{"level":"info","time":"2025-05-14T01:36:03Z","message":"Tune: setGCPercent to 100"}
2025-05-14T01:36:03Z INF cmd/search/main.go:24 > Reading config...
2025-05-14T01:36:03Z INF cmd/search/main.go:40 > Read target port: 8082
2025-05-14T01:36:03Z INF cmd/search/main.go:41 > Read consul address: consul:8500
2025-05-14T01:36:03Z INF cmd/search/main.go:42 > Read jaeger address: jaeger:6831
2025-05-14T01:36:03Z INF cmd/search/main.go:51 > Initializing jaeger agent [service name: search | host: jaeger:6831]...
2025-05-14T01:36:03Z INF tracing/tracer.go:26 > Jaeger client: adjusted sample ratio 1.000000
2025-05-14T01:36:03Z INF cmd/search/main.go:56 > Jaeger agent initialized
2025-05-14T01:36:03Z INF cmd/search/main.go:58 > Initializing consul agent [host: consul:8500]...
2025-05-14T01:36:03Z INF cmd/search/main.go:63 > Consul agent initialized
2025-05-14T01:36:03Z INF cmd/search/main.go:74 > Starting server...
2025-05-14T01:36:03Z INF registry/registry.go:91 > Trying to register service [ name: srv-search, id: 254379a6-bcf4-478c-a828-268a443478fb, address: 10.244.2.21:8082 ]
2025-05-14T01:36:03Z INF services/search/server.go:100 > Successfully registered in consul

Please take the next action
\end{agenttool}

\begin{agentthought}{Mitigation}
The logs for the "search" service do not indicate any errors or anomalies. The service seems to be initializing and registering successfully. Next, I will fetch the logs for the "frontend" service in the "test-hotel-reservation" namespace to investigate the "recommendation" operation anomaly.

\vspace{1em} \hrule \vspace{1em}

Call get\_logs with \{"namespace": "test-hotel-reservation", "service": "frontend"\}
\end{agentthought}

\begin{agenttool}{get\_logs}{Mitigation}
{"level":"info","time":"2025-05-14T01:35:46Z","message":"TLS disabled."}
{"level":"info","time":"2025-05-14T01:35:46Z","message":"Set global log level: info"}
{"level":"info","time":"2025-05-14T01:35:46Z","message":"Tune: setGCPercent to 100"}
2025-05-14T01:35:46Z INF cmd/frontend/main.go:23 > Reading config...
...
2025/05/14 01:35:46 grpc/lb/consul: error retrieving instances from Consul: Get "http://consul:8500/v1/health/service/srv-search?passing=1": dial tcp 10.96.73.138:8500: connect: connection refused
...
Please take the next action
\end{agenttool}

\begin{agentthought}{Mitigation}
The logs for the "frontend" service indicate a connection issue with the Consul service. The error message "dial tcp 10.96.73.138:8500: connect: connection refused" suggests that the frontend service is unable to connect to the Consul service, which is likely causing the anomaly in the "recommendation" operation.

To address this issue, I will check the status of the Consul service in the "test-hotel-reservation" namespace to ensure it is running and accessible.

\vspace{1em} \hrule \vspace{1em}

Call NL2Kubectl Tool with \{"nl\_query": "get pods -n test-hotel-reservation"\}
\end{agentthought}

\begin{agenttool}{NL2Kubectl Tool}{Mitigation}
NAME                                      READY   STATUS    RESTARTS      AGE
consul-b84877fb4-v9sgv                    1/1     Running   0             74s
frontend-69d95967fb-cgrjc                 1/1     Running   0             73s
geo-99d8c58c-5wrbn                        0/1     Error     2 (32s ago)   73s
jaeger-67445b6dd6-7kbpd                   1/1     Running   0             71s
memcached-profile-66bb9c8b9d-86bpf        1/1     Running   0             70s
memcached-rate-68585796cb-gtpg6           1/1     Running   0             67s
memcached-reserve-646c7dcd6b-vwz8t        1/1     Running   0             62s
mongodb-geo-5ff578bcb8-bg6cr              0/1     Pending   0             71s
mongodb-profile-758cb77f9f-thlpl          0/1     Pending   0             70s
mongodb-rate-56cc8659c9-7dwkw             0/1     Pending   0             66s
mongodb-recommendation-55699465f7-f9hng   0/1     Pending   0             64s
mongodb-reservation-5994859869-zvvmt      0/1     Pending   0             60s
mongodb-user-969c6c449-86tnh              0/1     Pending   0             58s
profile-676fcc86d5-v7c55                  1/1     Running   2 (26s ago)   69s
rate-c9bc58c85-klcq7                      1/1     Running   2 (24s ago)   65s
recommendation-8675c6c9ff-hqbl8           1/1     Running   2 (22s ago)   64s
reservation-7699b8fdd4-756lh              0/1     Error     1 (31s ago)   60s
search-65df96d8df-6lnw2                   1/1     Running   0             59s
user-8477d787d8-d8d45                     0/1     Error     1 (29s ago)   57s
\end{agenttool}

\begin{agentthought}{Mitigation}
The Consul pod is running, which suggests that the issue might not be with the Consul service itself but rather with network connectivity or configuration. The error message "connection refused" typically indicates that the service is not reachable at the specified address or port.

Given that the Consul pod is running, I will now check the service configuration for Consul to ensure it is correctly set up to expose the necessary ports.

\vspace{1em} \hrule \vspace{1em}

Call NL2Kubectl Tool with \{"nl\_query": "describe service consul -n test-hotel-reservation"\}
\end{agentthought}

\begin{agenttool}{NL2Kubectl Tool}{Mitigation}
Name:                     consul
...
Port:                     8300  8300/TCP
TargetPort:               8300/TCP
Endpoints:                10.244.2.17:8300
Port:                     8400  8400/TCP
TargetPort:               8400/TCP
Endpoints:                10.244.2.17:8400
Port:                     8500  8500/TCP
TargetPort:               8500/TCP
Endpoints:                10.244.2.17:8500
Port:                     8600  8600/UDP
TargetPort:               53/UDP
Endpoints:                10.244.2.17:53
...
\end{agenttool}

\begin{agentthought}{Mitigation}
The Consul service is correctly configured to expose port 8500, which is the port the frontend service is trying to connect to. The IP address and port are correctly set, and the endpoints are available. This suggests that the issue might be related to network policies or firewall rules that are preventing the frontend service from connecting to the Consul service.

To further investigate, I will check if there are any network policies in place that might be affecting connectivity between the frontend and Consul services.

\vspace{1em} \hrule \vspace{1em}

Call NL2Kubectl Tool with \{"nl\_query": "get networkpolicy -n test-hotel-reservation"\}
\end{agentthought}

\begin{agenttool}{NL2Kubectl Tool}{Mitigation}
\end{agenttool}

\begin{agentthought}{Mitigation}
There are no network policies listed in the "test-hotel-reservation" namespace, which means there are no restrictions on network traffic between the services. This suggests that the issue might be related to the configuration of the frontend service itself or a misconfiguration in the Consul service.

Given the information gathered, the root cause of the anomaly in the "frontend" service's "recommendation" operation appears to be a connectivity issue with the Consul service. The error message "connection refused" indicates that the frontend service is unable to reach the Consul service at the specified address and port.

Mitigation Plan:

1. Verify the network configuration and ensure that the frontend service can reach the Consul service on port 8500.

2. Check the firewall settings and ensure that there are no rules blocking traffic between the frontend and Consul services.

3. Review the configuration of the frontend service to ensure it is correctly set up to connect to the Consul service.

4. If necessary, restart the Consul service to ensure it is running correctly and accepting connections.

Thought: I have identified the root cause of the anomaly and outlined a mitigation plan. I will now submit the solution.

\vspace{1em} \hrule \vspace{1em}

Call submit with \{\}
\end{agentthought}

\begin{frameworklog}{Submission Validation}{1}
Submission triggered. Validating...
Waiting 15 seconds for changes to take effect...
Testing workload generator...
...
  117 requests in 10.00s, 30.62KB read
  Non-2xx or 3xx responses: 117
Requests/sec:     11.70  
Transfer/sec:      3.06KB
\end{frameworklog}

\begin{agenttool}{submit}{Mitigation}
The system is not in a valid state. Please resolve all issues before submitting.
\end{agenttool}

\begin{agentthought}{Mitigation}
The submission failed because the system is not in a valid state, indicating unresolved issues. I need to further investigate and resolve any remaining issues before submitting the solution.

The "geo" operation in the "search" service also has an anomaly. I will check the logs for the "geo" pod to gather more information about the error status.

\vspace{1em} \hrule \vspace{1em}

Call get\_logs with \{"namespace": "test-hotel-reservation", "service": "geo"\}
\end{agentthought}

\begin{agenttool}{get\_logs}{Mitigation}
...
2025-05-14T01:37:15Z INF cmd/geo/main.go:37 > Initializing DB connection...
2025-05-14T01:37:27Z PNC cmd/geo/db.go:26 > no reachable servers
panic: no reachable servers

goroutine 1 [running]:
github.com/rs/zerolog.(*Logger).Panic.func1({0x96e211, 0x0})
...
Please take the next action
\end{agenttool}

\begin{agentthought}{Mitigation}
The logs for the "geo" service indicate a panic due to "no reachable servers," which suggests a connectivity issue with the MongoDB database. The error occurs during the initialization of the database connection, indicating that the "geo" service is unable to connect to the MongoDB instance at "mongodb-geo:27017."

To address this issue, I will check the status of the MongoDB pod for the "geo" service to ensure it is running and accessible.

\vspace{1em} \hrule \vspace{1em}

Call NL2Kubectl Tool with \{"nl\_query": "describe pod mongodb-geo-5ff578bcb8-bg6cr -n test-hotel-reservation"\}
\end{agentthought}

\begin{agenttool}{NL2Kubectl Tool}{Mitigation}
Name:             mongodb-geo-5ff578bcb8-bg6cr
Namespace:        test-hotel-reservation
Priority:         0
Service Account:  default
Node:             <none>
Labels:           io.kompose.service=mongodb-geo
                  pod-template-hash=5ff578bcb8
...
Events:
  Type     Reason            Age   From               Message
  ----     ------            ----  ----               -------
  Warning  FailedScheduling  2m    default-scheduler  0/3 nodes are available: pod has unbound immediate PersistentVolumeClaims. preemption: 0/3 nodes are available: 3 Preemption is not helpful for scheduling.
\end{agenttool}

\begin{agentthought}{Mitigation}
The "mongodb-geo" pod is in a "Pending" state due to a failed scheduling event. The message "pod has unbound immediate PersistentVolumeClaims" indicates that the pod is unable to start because it cannot bind to the required PersistentVolumeClaim (PVC) named "geo-pvc."

To resolve this issue, I need to check the status of the "geo-pvc" to ensure it is correctly configured and available.

\vspace{1em} \hrule \vspace{1em}

Call NL2Kubectl Tool with \{"nl\_query": "describe pvc geo-pvc -n test-hotel-reservation"\}
\end{agentthought}

\begin{agenttool}{NL2Kubectl Tool}{Mitigation}
Name:          geo-pvc
Namespace:     test-hotel-reservation
StorageClass:  geo-storage
Status:        Pending
Volume:        
Labels:        <none>
Annotations:   <none>
Finalizers:    [kubernetes.io/pvc-protection]
Capacity:      
Access Modes:  
VolumeMode:    Filesystem
Used By:       mongodb-geo-5ff578bcb8-bg6cr
Events:
  Type     Reason              Age                From                         Message
  ----     ------              ----               ----                         -------
  Warning  ProvisioningFailed  4s (x9 over 2m4s)  persistentvolume-controller  storageclass.storage.k8s.io "geo-storage" not found
\end{agenttool}

\begin{agentthought}{Mitigation}
The "geo-pvc" is in a "Pending" state because the storage class "geo-storage" is not found. This is preventing the PersistentVolumeClaim from being bound, which in turn is causing the "mongodb-geo" pod to remain in a "Pending" state.

To resolve this issue, I need to create or correct the storage class "geo-storage" so that the PVC can be bound and the pod can start.

Mitigation Plan:

1. Create a storage class named "geo-storage" with the appropriate configuration to support the "geo-pvc" PersistentVolumeClaim.

2. Ensure that the storage class is available and correctly configured to allow the PVC to bind.

3. Once the storage class is created, the "geo-pvc" should bind successfully, allowing the "mongodb-geo" pod to start.

I will now proceed to create the storage class "geo-storage."

\vspace{1em} \hrule \vspace{1em}

Call NL2Kubectl Tool with \{"nl\_query": "create storageclass geo-storage -{}-provisioner=kubernetes.io/aws-ebs -{}-parameters type=gp2"\}
\end{agentthought}

\begin{agenttool}{NL2Kubectl Tool}{Mitigation}
Dry-run failed. Potentially it's an invalid command. stderr: error: unknown flag: --provisioner
See 'kubectl create --help' for usage.
\end{agenttool}

\begin{agentthought}{Mitigation}
It seems the command to create the storage class was incorrect. I will correct the command and attempt to create the storage class "geo-storage" using a valid configuration.

\vspace{1em} \hrule \vspace{1em}

Call NL2Kubectl Tool with \{"nl\_query": "apply -f - <<EOF{\textbackslash}napiVersion: storage.k8s.io/v1{\textbackslash}nkind: StorageClass{\textbackslash}nmetadata:{\textbackslash}n  name: geo-storage{\textbackslash}nprovisioner: kubernetes.io/aws-ebs{\textbackslash}nparameters:{\textbackslash}n  type: gp2{\textbackslash}nEOF"\}
\end{agentthought}

\begin{agenttool}{NL2Kubectl Tool}{Mitigation}
storageclass.storage.k8s.io/geo-storage created
\end{agenttool}

\begin{agentthought}{Mitigation}
The storage class "geo-storage" has been successfully created. I will now check the status of the "geo-pvc" to ensure it is bound and the "mongodb-geo" pod can start.

\vspace{1em} \hrule \vspace{1em}

Call NL2Kubectl Tool with \{"nl\_query": "describe pvc geo-pvc -n test-hotel-reservation"\}
\end{agentthought}

\begin{agenttool}{NL2Kubectl Tool}{Mitigation}
Name:          geo-pvc
Namespace:     test-hotel-reservation
StorageClass:  geo-storage
Status:        Pending
Volume:        
Labels:        <none>
Annotations:   <none>
Finalizers:    [kubernetes.io/pvc-protection]
Capacity:      
Access Modes:  
VolumeMode:    Filesystem
Used By:       mongodb-geo-5ff578bcb8-bg6cr
Events:
  Type     Reason              Age                  From                         Message
  ----     ------              ----                 ----                         -------
  Warning  ProvisioningFailed  14s (x9 over 2m14s)  persistentvolume-controller  storageclass.storage.k8s.io "geo-storage" not found
\end{agenttool}

\begin{agentthought}{Mitigation}
The "geo-pvc" is still in a "Pending" state, and the error message indicates that the storage class "geo-storage" is not found. This suggests that there might be an issue with the creation of the storage class or its recognition by the persistent volume controller.

To resolve this, I will verify the existence of the storage class "geo-storage" and ensure it is correctly configured.

\vspace{1em} \hrule \vspace{1em}

Call NL2Kubectl Tool with \{"nl\_query": "get storageclass geo-storage"\}
\end{agentthought}

\begin{agenttool}{NL2Kubectl Tool}{Mitigation}
NAME          PROVISIONER             RECLAIMPOLICY   VOLUMEBINDINGMODE   ALLOWVOLUMEEXPANSION   AGE
geo-storage   kubernetes.io/aws-ebs   Delete          Immediate           false                  6s
\end{agenttool}

\begin{agentthought}{Mitigation}
The storage class "geo-storage" exists and is correctly configured with the provisioner "kubernetes.io/aws-ebs." However, the persistent volume controller is still unable to find it, which suggests a potential issue with the cluster configuration or the controller itself.

Given the information gathered, the root cause of the anomaly in the "geo" operation of the "search" service is the inability to bind the "geo-pvc" due to the persistent volume controller not recognizing the storage class "geo-storage."

Mitigation Plan:

1. Verify the cluster configuration to ensure the persistent volume controller is correctly set up to recognize storage classes.

2. Check the logs of the persistent volume controller for any errors or misconfigurations.

3. If necessary, restart the persistent volume controller to ensure it is running correctly and can recognize the storage class.

Thought: I have identified the root cause of the anomaly and outlined a mitigation plan. I will now submit the solution.

\vspace{1em} \hrule \vspace{1em}

Call submit with \{\}
\end{agentthought}

\begin{frameworklog}{Submission Validation}{1}
Submission triggered. Validating...
Waiting 15 seconds for changes to take effect...
Testing workload generator...
...
  117 requests in 10.00s, 30.62KB read
  Non-2xx or 3xx responses: 117
Requests/sec:     11.70  
Transfer/sec:      3.06KB
\end{frameworklog}

\begin{agenttool}{submit}{Mitigation}
The system is not in a valid state. Please resolve all issues before submitting.
\end{agenttool}

\begin{tcolorbox}[frame hidden]
    {$\cdots$} \\
    {\small\ttfamily A series of observation and submission.} \\
    {$\cdots$}
\end{tcolorbox}

\begin{agentanswer}{Mitigation}{Mitigation}
Root Cause Analysis:
1. The "frontend" service's "recommendation" operation anomaly is due to a connectivity issue with the Consul service. The error message "connection refused" indicates that the frontend service is unable to reach the Consul service at the specified address and port.
2. The "geo" operation in the "search" service is experiencing a connectivity issue with the MongoDB database. The error "no reachable servers" suggests that the "geo" service is unable to connect to the MongoDB instance due to the "geo-pvc" being in a "Pending" state.
3. The "geo-pvc" is unable to bind because the persistent volume controller is not recognizing the storage class "geo-storage," despite it being correctly configured.

Mitigation Plan:
1. For the "frontend" service:
   - Verify the network configuration and ensure that the frontend service can reach the Consul service on port 8500.
   - Check firewall settings to ensure there are no rules blocking traffic between the frontend and Consul services.
   - Review the configuration of the frontend service to ensure it is correctly set up to connect to the Consul service.
   - Restart the Consul service if necessary to ensure it is running correctly and accepting connections.

2. For the "geo" operation:
   - Verify the cluster configuration to ensure the persistent volume controller is correctly set up to recognize storage classes.
   - Check the cluster setup to ensure all necessary components are installed and configured.
   - Restart the persistent volume controller if necessary to ensure it is running correctly and can recognize the storage class "geo-storage."

These steps should help resolve the anomalies in the "frontend" and "search" services.
\end{agentanswer}

\begin{frameworklog}{Validation \& Retry}{1}
-----------------------------------------------------------------------
  117 requests in 10.00s, 44.57KB read
  Non-2xx or 3xx responses: 63
Requests/sec:     11.70  
Transfer/sec:      4.46KB

Adding issues ['  Non-2xx or 3xx responses: 117']
Issues found with pod geo-99d8c58c-5wrbn:
  - Container hotel-reserv-geo is in CrashLoopBackOff
...
Issues found with pod user-8477d787d8-d8d45:
  - Container hotel-reserv-user is in CrashLoopBackOff
Found 12 issues in the cluster.
Adding issues ['Container hotel-reserv-geo is in CrashLoopBackOff', ...]
Validation result: {'success': False, 'issues': ['  Non-2xx or 3xx responses: 117', 'Container hotel-reserv-geo is in CrashLoopBackOff', ...]}
Output written to: eval/05-13_21-28-46-redeploy_without_PV-mitigation-1/lumyn_output/agent_output_0.json
Analyzing previous run results for improvement...
Reflection generated successfully.
\end{frameworklog}

\begin{tcolorbox}[frame hidden, title={Note (not part of the trajectory)}]
  \scriptsize A stroageclass has been created but not deleted in the first run.
  However it's created using the wrong provisioner.
  The Persistent Volume Controller will report the following error:
  \vspace{0.5em}

  {\ttfamily\color{red!70!black}
  -- Waiting for a volume to be created either by the external provisioner 'ebs.csi.aws.com' or manually by the system administrator. If volume creation is delayed, please verify that the provisioner is running and correctly registered.
  }

  \vspace{1em}

  Any overwrite to such existing stroageclass will result in errors like:
  \vspace{0.5em}

  {\ttfamily\color{red!70!black}
  The StorageClass "<storage-class-name>" is invalid: 

  * provisioner: Forbidden: updates to provisioner are forbidden.

  * volumeBindingMode: Invalid value: "WaitForFirstConsumer": field is immutable
  }

  \vspace{1em}

  Rollback prevents such overwrites and errors in the next run.
\end{tcolorbox}

\begin{agenttask}{Rollback}{Rollback}
You are given a system and a rollback tool. The system is either in a initial faulty state, or in the state you left it incompletely repaired after the previous round of mitigation. You need to rollback the system to the initial state since the operations you performed in the previous round may negatively impact your current round of mitigation.
You should run the rollback tool as many times as you can, since this rollback tool undos the last operation performed on the system. You should run the rollback tool until it returns that there are no more operations to undo.
\end{agenttask}

\begin{agentthought}{Rollback}
To rollback the system to its initial state, I need to use the rollback tool repeatedly until it indicates that there are no more operations to undo.

\vspace{1em} \hrule \vspace{1em}

Call rollback\_tool with \{"argument": "rollback the last operation"\}
\end{agentthought}

\begin{agenttool}{rollback\_tool}{Rollback}
Rolled back the previous command: kubectl apply -f - <<EOF
apiVersion: storage.k8s.io/v1
kind: StorageClass
metadata:
  name: geo-storage
provisioner: kubernetes.io/aws-ebs
parameters:
  type: gp2
EOF, using rollback:kubectl delete storageclass.storage.k8s.io geo-storage -n test-hotel-reservation
\end{agenttool}

\begin{agentthought}{Rollback}
Thought: I need to provide a different argument to the rollback tool to continue the rollback process.

\vspace{1em} \hrule \vspace{1em}

Call rollback\_tool with \{"argument": "undo the previous operation"\}
\end{agentthought}

\begin{agenttool}{rollback\_tool}{Rollback}
No more actions to rollback.
\end{agenttool}

\begin{agentanswer}{Rollback}{Rollback}
None.
\end{agentanswer}

\begin{agenttask}{Bootstrapping}{Diagnosis}
Analyze the traces to localize which part of the system, e.g., a service, a component, a pod, a node, etc., has anomalies. Always use the get_traces tool first; then use a read_traces tool to help you with the localization. Note that it may not be helpful to use the tools too many times. You are working on the namespace 'test-hotel-reservation', please use this namespace when using the tools.
\end{agenttask}

\begin{agentthought}{Diagnosis}
I need to fetch traces from the 'test-hotel-reservation' namespace to analyze and identify any anomalies in the system. I'll start by using the get\_traces tool to retrieve the necessary data.

\vspace{1em} \hrule \vspace{1em}

Call get\_traces with \{"namespace": "test-hotel-reservation", "duration": 60\}
\end{agentthought}

\begin{agenttool}{get\_traces}{Diagnosis}
Traces data exported to: /eval/xagent/trace_output/traces_1747186964.csv
Please take the next action
\end{agenttool}

\begin{agentthought}{Diagnosis}
Thought: I have successfully fetched the traces data. Now, I need to read the traces from the file to analyze and identify any anomalies in the system.

\vspace{1em} \hrule \vspace{1em}

Call read\_traces with \{"file\_path": "/eval/xagent/trace\_output/traces\_1747186964.csv"\}
\end{agentthought}

\begin{agenttool}{read\_traces}{Diagnosis}
The following list of services are the last service the request reached before returning an error, these services may not be the exact place of the root cause, so you may need to dig deeper. These services correspond to a list of operations happened there, which is the service relying on and could be where the root cause comes from. Here is the list:
1: {service: "frontend", operation: "on/GetRecommendations"}
2: {service: "search", operation: "geo"}
3: {service: "frontend", operation: "user"}
\end{agenttool}

\begin{agentanswer}{Bootstrapping}{Diagnosis}
- List of entities with anomalies:
  1. Service: "frontend", Operation: "on/GetRecommendations"
  2. Service: "search", Operation: "geo"
  3. Service: "frontend", Operation: "user"

- Possible root cause entities:
  - The anomalies may originate from the "frontend" service, particularly in the "on/GetRecommendations" and "user" operations, as well as the "search" service in the "geo" operation. Further analysis can start from these root cause entities.
\end{agentanswer}

\begin{agenttask}{Mitigation}{Mitigation}
The problem description is '''You are an expert DevOps engineer assigned to mitigate anomalies in a deployed service...

Reflection on the previous run: Root Cause:
1. The anomaly in the "frontend" service's "recommendation" operation is due to a connectivity issue with the Consul service, indicated by the error message "connection refused."
2. The anomaly in the "geo" operation of the "search" service is due to the inability to bind the "geo-pvc" because the persistent volume controller is not recognizing the storage class "geo-storage."

Mitigation Plan:
1. For the "frontend" service:
   - Verify the network configuration and ensure that the frontend service can reach the Consul service on port 8500.
   - Check the firewall settings and ensure that there are no rules blocking traffic between the frontend and Consul services.
   - Review the configuration of the frontend service to ensure it is correctly set up to connect to the Consul service.
   - If necessary, restart the Consul service to ensure it is running correctly and accepting connections.

2. For the "geo" operation in the "search" service:
   - Verify the cluster configuration to ensure the persistent volume controller is correctly set up to recognize storage classes.
   - Check the cluster setup to ensure all necessary components are installed and configured.
   - If necessary, restart the persistent volume controller to ensure it is running correctly and can recognize the storage class..


The following issues were found in the cluster:
-   Non-2xx or 3xx responses: 117
- Container hotel-reserv-geo is in CrashLoopBackOff
- Pod mongodb-geo-5ff578bcb8-bg6cr is in Pending state
- Pod mongodb-profile-758cb77f9f-thlpl is in Pending state
- ...
\end{agenttask}

\begin{agentthought}{Mitigation}
The previous run has already identified the root causes and provided a mitigation plan. I will follow the plan to resolve the issues. The plan involves two main tasks: fixing the connectivity issue with the Consul service for the "frontend" service and addressing the persistent volume controller issue for the "geo" operation in the "search" service.

First, I will verify the network configuration and ensure that the frontend service can reach the Consul service on port 8500. I will check the firewall settings and review the configuration of the frontend service. If necessary, I will restart the Consul service.

Next, I will verify the cluster configuration to ensure the persistent volume controller is correctly set up to recognize storage classes. I will check the cluster setup and restart the persistent volume controller if necessary.

Let's start by checking the network configuration for the "frontend" service to ensure it can reach the Consul service.

\vspace{1em} \hrule \vspace{1em}

Call NL2Kubectl Tool with \{"nl\_query": "describe service frontend -n test-hotel-reservation"\}
\end{agentthought}

\begin{agenttool}{NL2Kubectl Tool}{Mitigation}
Name:                     frontend
Namespace:                test-hotel-reservation
Labels:                   io.kompose.service=frontend
Annotations:              kompose.cmd: kompose convert
                          kompose.version: 1.22.0 (955b78124)
Selector:                 io.kompose.service=frontend
Type:                     ClusterIP
IP Family Policy:         SingleStack
IP Families:              IPv4
IP:                       10.96.54.225
IPs:                      10.96.54.225
Port:                     5000  5000/TCP
TargetPort:               5000/TCP
...
\end{agenttool}

\begin{agentthought}{Mitigation}
The "frontend" service is configured with a ClusterIP and is listening on port 5000. However, the issue is related to connectivity with the Consul service on port 8500. I need to check if the Consul service is running and accessible on the expected port.

Next, I will describe the Consul service to verify its configuration and ensure it is running correctly.

\vspace{1em} \hrule \vspace{1em}

Call NL2Kubectl Tool with \{"nl\_query": "describe service consul -n test-hotel-reservation"\}
\end{agentthought}

\begin{agenttool}{NL2Kubectl Tool}{Mitigation}
Name:                     consul
...
Port:                     8300  8300/TCP
TargetPort:               8300/TCP
Endpoints:                10.244.2.17:8300
Port:                     8400  8400/TCP
TargetPort:               8400/TCP
Endpoints:                10.244.2.17:8400
Port:                     8500  8500/TCP
TargetPort:               8500/TCP
Endpoints:                10.244.2.17:8500
Port:                     8600  8600/UDP
TargetPort:               53/UDP
Endpoints:                10.244.2.17:53
...
\end{agenttool}

\begin{agentthought}{Mitigation}
The Consul service is correctly configured with a ClusterIP and is listening on port 8500, which matches the expected configuration. The endpoints are also correctly set up. This suggests that the issue might be related to firewall settings or the frontend service configuration.

I will now check the firewall settings to ensure there are no rules blocking traffic between the frontend and Consul services.

\vspace{1em} \hrule \vspace{1em}

Call NL2Kubectl Tool with \{"nl\_query": "get networkpolicy -n test-hotel-reservation"\}
\end{agentthought}

\begin{agenttool}{NL2Kubectl Tool}{Mitigation}
\end{agenttool}

\begin{agentthought}{Mitigation}
There are no network policies defined in the "test-hotel-reservation" namespace, which means there are no firewall rules blocking traffic between the frontend and Consul services. The issue might be related to the frontend service configuration.

I will now check the configuration of the frontend service to ensure it is correctly set up to connect to the Consul service.

\vspace{1em} \hrule \vspace{1em}

Call NL2Kubectl Tool with \{"nl\_query": "describe deployment frontend -n test-hotel-reservation"\}
\end{agentthought}

\begin{agenttool}{NL2Kubectl Tool}{Mitigation}
Name:                   frontend
Namespace:              test-hotel-reservation
CreationTimestamp:      Tue, 13 May 2025 21:35:37 -0400
Labels:                 io.kompose.service=frontend
Annotations:            deployment.kubernetes.io/revision: 1
                        kompose.cmd: kompose convert
                        kompose.version: 1.22.0 (955b78124)
Selector:               io.kompose.service=frontend
Replicas:               1 desired | 1 updated | 1 total | 1 available | 0 unavailable
StrategyType:           RollingUpdate
...
\end{agenttool}

\begin{agentthought}{Mitigation}
The frontend deployment is configured correctly, and there are no apparent issues with the environment variables or command setup. The deployment is running and available, which suggests that the connectivity issue might be transient or related to the Consul service itself.

I will proceed to restart the Consul service to ensure it is running correctly and accepting connections.

\vspace{1em} \hrule \vspace{1em}

Call NL2Kubectl Tool with \{"nl\_query": "rollout restart deployment consul -n test-hotel-reservation"\}
\end{agentthought}

\begin{agenttool}{NL2Kubectl Tool}{Mitigation}
deployment.apps/consul restarted
\end{agenttool}

\begin{agentthought}{Mitigation}
The Consul deployment has been restarted. This should help resolve any transient connectivity issues between the frontend service and the Consul service.

Next, I will address the issue with the "geo" operation in the "search" service, which is related to the persistent volume controller not recognizing the storage class "geo-storage." I will start by checking the persistent volume claims to verify their status.

\vspace{1em} \hrule \vspace{1em}

Call NL2Kubectl Tool with \{"nl\_query": "get pvc -n test-hotel-reservation"\}
\end{agentthought}

\begin{agenttool}{NL2Kubectl Tool}{Mitigation}
NAME                 STATUS    VOLUME   CAPACITY   ACCESS MODES   STORAGECLASS             VOLUMEATTRIBUTESCLASS   AGE
geo-pvc              Pending                                      geo-storage              <unset>                 7m23s
profile-pvc          Pending                                      profile-storage          <unset>                 7m20s
rate-pvc             Pending                                      rate-storage             <unset>                 7m16s
recommendation-pvc   Pending                                      recommendation-storage   <unset>                 7m14s
reservation-pvc      Pending                                      reservation-storage      <unset>                 7m11s
user-pvc             Pending                                      user-storage             <unset>                 7m8s
\end{agenttool}

\begin{agentthought}{Mitigation}
The persistent volume claims (PVCs) are all in a "Pending" state, indicating that they are not being bound to any persistent volumes. This is likely due to the storage classes not being recognized by the persistent volume controller.

I will now check the storage classes to ensure they are correctly configured and recognized by the cluster.

\vspace{1em} \hrule \vspace{1em}

Call NL2Kubectl Tool with \{"nl\_query": "get storageclass"\}
\end{agentthought}

\begin{agenttool}{NL2Kubectl Tool}{Mitigation}
NAME                         PROVISIONER             RECLAIMPOLICY   VOLUMEBINDINGMODE      ALLOWVOLUMEEXPANSION   AGE
openebs-device               openebs.io/local        Delete          WaitForFirstConsumer   false                  13m
openebs-hostpath (default)   openebs.io/local        Delete          WaitForFirstConsumer   false                  13m
standard (default)           rancher.io/local-path   Delete          WaitForFirstConsumer   false                  14m
\end{agenttool}

\begin{agentthought}{Mitigation}
The storage classes "geo-storage", "profile-storage", "rate-storage", "recommendation-storage", "reservation-storage", and "user-storage" are not listed among the available storage classes. This indicates that these storage classes are not defined in the cluster, which is why the PVCs are in a "Pending" state.

To resolve this issue, I need to create the missing storage classes. I will start by creating the "geo-storage" class.

\vspace{1em} \hrule \vspace{1em}

Call NL2Kubectl Tool with \{"nl\_query": "create storageclass geo-storage -{}-provisioner=rancher.io/local-path -{}-reclaim-policy=Delete -{}-volume-binding-mode=WaitForFirstConsumer"\}
\end{agentthought}

\begin{agenttool}{NL2Kubectl Tool}{Mitigation}
Dry-run failed. Potentially it's an invalid command. stderr: error: unknown flag: --provisioner
See 'kubectl create --help' for usage.
\end{agenttool}

\begin{agentgoldthought}{1}{Mitigation}
It seems the command to create a storage class was incorrect. I will use the correct syntax to create the "geo-storage" storage class using a YAML file.

\vspace{1em} \hrule \vspace{1em}

Call NL2Kubectl Tool with \{"nl\_query": "apply -f - <<EOF{\textbackslash}napiVersion: storage.k8s.io/v1{\textbackslash}nkind: StorageClass{\textbackslash}nmetadata:{\textbackslash}n  name: geo-storage{\textbackslash}nprovisioner: rancher.io/local-path{\textbackslash}nreclaimPolicy: Delete{\textbackslash}nvolumeBindingMode: WaitForFirstConsumer{\textbackslash}nEOF"\}
\end{agentgoldthought}

\begin{agenttool}{NL2Kubectl Tool}{Mitigation}
storageclass.storage.k8s.io/geo-storage created
\end{agenttool}

\begin{agentgoldthought}{2}{Mitigation}
The "geo-storage" storage class has been successfully created. I will proceed to create the other missing storage classes: "profile-storage", "rate-storage", "recommendation-storage", "reservation-storage", and "user-storage".

\vspace{1em} \hrule \vspace{1em}

Call NL2Kubectl Tool with \{"nl\_query": "apply -f - <<EOF{\textbackslash}napiVersion: storage.k8s.io/v1{\textbackslash}nkind: StorageClass{\textbackslash}nmetadata:{\textbackslash}n  name: profile-storage{\textbackslash}nprovisioner: rancher.io/local-path{\textbackslash}nreclaimPolicy: Delete{\textbackslash}nvolumeBindingMode: WaitForFirstConsumer{\textbackslash}nEOF"\}
\end{agentgoldthought}

\begin{agenttool}{NL2Kubectl Tool}{Mitigation}
storageclass.storage.k8s.io/profile-storage created
\end{agenttool}

\begin{agentgoldthought}{3}{Mitigation}
The "profile-storage" storage class has been successfully created. I will continue to create the remaining storage classes.

\vspace{1em} \hrule \vspace{1em}

Call NL2Kubectl Tool with \{"nl\_query": "apply -f - <<EOF{\textbackslash}napiVersion: storage.k8s.io/v1{\textbackslash}nkind: StorageClass{\textbackslash}nmetadata:{\textbackslash}n  name: rate-storage{\textbackslash}nprovisioner: rancher.io/local-path{\textbackslash}nreclaimPolicy: Delete{\textbackslash}nvolumeBindingMode: WaitForFirstConsumer{\textbackslash}nEOF"\}
\end{agentgoldthought}

\begin{agenttool}{NL2Kubectl Tool}{Mitigation}
storageclass.storage.k8s.io/rate-storage created
\end{agenttool}

\begin{agentgoldthought}{4}{Mitigation}
The "rate-storage" storage class has been successfully created. I will proceed to create the "recommendation-storage" storage class.

\vspace{1em} \hrule \vspace{1em}

Call NL2Kubectl Tool with \{"nl\_query": "apply -f - <<EOF{\textbackslash}napiVersion: storage.k8s.io/v1{\textbackslash}nkind: StorageClass{\textbackslash}nmetadata:{\textbackslash}n  name: recommendation-storage{\textbackslash}nprovisioner: rancher.io/local-path{\textbackslash}nreclaimPolicy: Delete{\textbackslash}nvolumeBindingMode: WaitForFirstConsumer{\textbackslash}nEOF"\}
\end{agentgoldthought}

\begin{agenttool}{NL2Kubectl Tool}{Mitigation}
storageclass.storage.k8s.io/recommendation-storage created
\end{agenttool}

\begin{agentgoldthought}{5}{Mitigation}
The "recommendation-storage" storage class has been successfully created. I will proceed to create the "reservation-storage" storage class.

\vspace{1em} \hrule \vspace{1em}

Call NL2Kubectl Tool with \{"nl\_query": "apply -f - <<EOF{\textbackslash}napiVersion: storage.k8s.io/v1{\textbackslash}nkind: StorageClass{\textbackslash}nmetadata:{\textbackslash}n  name: reservation-storage{\textbackslash}nprovisioner: rancher.io/local-path{\textbackslash}nreclaimPolicy: Delete{\textbackslash}nvolumeBindingMode: WaitForFirstConsumer{\textbackslash}nEOF"\}
\end{agentgoldthought}

\begin{agenttool}{NL2Kubectl Tool}{Mitigation}
storageclass.storage.k8s.io/reservation-storage created
\end{agenttool}

\begin{agentgoldthought}{6}{Mitigation}
The "reservation-storage" storage class has been successfully created. I will proceed to create the "user-storage" storage class.

\vspace{1em} \hrule \vspace{1em}

Call NL2Kubectl Tool with \{"nl\_query": "apply -f - <<EOF{\textbackslash}napiVersion: storage.k8s.io/v1{\textbackslash}nkind: StorageClass{\textbackslash}nmetadata:{\textbackslash}n  name: user-storage{\textbackslash}nprovisioner: rancher.io/local-path{\textbackslash}nreclaimPolicy: Delete{\textbackslash}nvolumeBindingMode: WaitForFirstConsumer{\textbackslash}nEOF"\}
\end{agentgoldthought}

\begin{agenttool}{NL2Kubectl Tool}{Mitigation}
storageclass.storage.k8s.io/user-storage created
\end{agenttool}

\begin{agentthought}{Mitigation}
The "user-storage" storage class has been successfully created. All the necessary storage classes are now in place. This should resolve the issue with the persistent volume claims being in a "Pending" state.

I will now check the status of the persistent volume claims to ensure they are bound correctly.

\vspace{1em} \hrule \vspace{1em}

Call NL2Kubectl Tool with \{"nl\_query": "get pvc -n test-hotel-reservation"\}
\end{agentthought}

\begin{agenttool}{NL2Kubectl Tool}{Mitigation}
NAME                 STATUS    VOLUME                                     CAPACITY   ACCESS MODES   STORAGECLASS             VOLUMEATTRIBUTESCLASS   AGE
geo-pvc              Bound     pvc-2937753a-cc60-4904-9371-d89b0db18ecd   1Gi        RWO            geo-storage              <unset>                 7m53s
profile-pvc          Bound     pvc-a4f1d897-89ca-4cdf-8f94-0958942aa243   1Gi        RWO            profile-storage          <unset>                 7m50s
rate-pvc             Pending                                                                        rate-storage             <unset>                 7m46s
recommendation-pvc   Pending                                                                        recommendation-storage   <unset>                 7m44s
reservation-pvc      Pending                                                                        reservation-storage      <unset>                 7m41s
user-pvc             Pending                                                                        user-storage             <unset>                 7m38s
\end{agenttool}

\begin{agentthought}{Mitigation}
The "geo-pvc" and "profile-pvc" are now bound, indicating that the storage classes are being recognized. However, the other PVCs are still in a "Pending" state. This might be due to resource constraints or scheduling issues.

I will check the events for one of the pending PVCs to understand why it is not being bound.

\vspace{1em} \hrule \vspace{1em}

Call NL2Kubectl Tool with \{"nl\_query": "describe pvc rate-pvc -n test-hotel-reservation"\}
\end{agentthought}

\begin{agenttool}{NL2Kubectl Tool}{Mitigation}
Name:          rate-pvc
Namespace:     test-hotel-reservation
StorageClass:  rate-storage
Status:        Pending
Volume:        
Labels:        <none>
Annotations:   volume.beta.kubernetes.io/storage-provisioner: rancher.io/local-path
               volume.kubernetes.io/selected-node: kind-worker2
               volume.kubernetes.io/storage-provisioner: rancher.io/local-path
Finalizers:    [kubernetes.io/pvc-protection]
Capacity:      
Access Modes:  
VolumeMode:    Filesystem
Used By:       mongodb-rate-56cc8659c9-7dwkw
Events:
  Type     Reason                Age                    From                                                                                                Message
  ----     ------                ----                   ----                                                                                                -------
  Warning  ProvisioningFailed    100s (x26 over 7m48s)  persistentvolume-controller                                                                         storageclass.storage.k8s.io "rate-storage" not found
  Normal   WaitForPodScheduled   10s                    persistentvolume-controller                                                                         waiting for pod mongodb-rate-56cc8659c9-7dwkw to be scheduled
  Normal   ExternalProvisioning  5s                     persistentvolume-controller                                                                         Waiting for a volume to be created either by the external provisioner 'rancher.io/local-path' or manually by the system administrator. If volume creation is delayed, please verify that the provisioner is running and correctly registered.
  Normal   Provisioning          5s                     rancher.io/local-path_local-path-provisioner-58cc7856b6-n29rr_bf374c2e-efd5-4d8b-a53f-0ac317a46e1a  External provisioner is provisioning volume for claim "test-hotel-reservation/rate-pvc"
\end{agenttool}

\begin{agentthought}{Mitigation}
The "rate-pvc" is pending due to the storage class "rate-storage" not being found initially, but it seems the provisioner is now attempting to provision the volume. This indicates that the issue might resolve itself as the provisioner completes its task.

I will check the status of the persistent volume claims again to see if they have been bound.

\vspace{1em} \hrule \vspace{1em}

Call NL2Kubectl Tool with \{"nl\_query": "get pvc -n test-hotel-reservation"\}
\end{agentthought}

\begin{agenttool}{NL2Kubectl Tool}{Mitigation}
NAME                 STATUS    VOLUME                                     CAPACITY   ACCESS MODES   STORAGECLASS             VOLUMEATTRIBUTESCLASS   AGE
geo-pvc              Bound     pvc-2937753a-cc60-4904-9371-d89b0db18ecd   1Gi        RWO            geo-storage              <unset>                 7m57s
profile-pvc          Bound     pvc-a4f1d897-89ca-4cdf-8f94-0958942aa243   1Gi        RWO            profile-storage          <unset>                 7m54s
rate-pvc             Pending                                                                        rate-storage             <unset>                 7m50s
recommendation-pvc   Pending                                                                        recommendation-storage   <unset>                 7m48s
reservation-pvc      Bound     pvc-b7dfe091-9243-41ba-8175-92d350c81f95   1Gi        RWO            reservation-storage      <unset>                 7m45s
user-pvc             Pending                                                                        user-storage             <unset>                 7m42s
\end{agenttool}

\begin{agentthought}{Mitigation}
The "reservation-pvc" has been successfully bound, but "rate-pvc", "recommendation-pvc", and "user-pvc" are still pending. The provisioner is working on these claims, and they may resolve shortly.

I will now check the logs of the "hotel-reserv-geo" container to see if there are any errors related to the CrashLoopBackOff state.

\vspace{1em} \hrule \vspace{1em}

Call get\_logs with \{"namespace": "test-hotel-reservation", "service": "hotel-reserv-geo"\}
\end{agentthought}

\begin{agenttool}{get\_logs}{Mitigation}
Error: Your service/namespace does not exist. Use kubectl to check.
Please take the next action
\end{agenttool}

\begin{agentthought}{Mitigation}
It seems there was an error retrieving logs for the "hotel-reserv-geo" container, possibly due to a misconfiguration or the service not existing. I will check the pods in the "test-hotel-reservation" namespace to verify their status and identify any issues.

\vspace{1em} \hrule \vspace{1em}

Call NL2Kubectl Tool with \{"nl\_query": "get pods -n test-hotel-reservation"\}
\end{agentthought}

\begin{agenttool}{NL2Kubectl Tool}{Mitigation}
NAME                                      READY   STATUS             RESTARTS        AGE
consul-65844d4555-5vpkm                   1/1     Running            0               40s
frontend-69d95967fb-cgrjc                 1/1     Running            0               8m3s
geo-99d8c58c-5wrbn                        0/1     CrashLoopBackOff   6 (41s ago)     8m3s
jaeger-67445b6dd6-7kbpd                   1/1     Running            0               8m1s
memcached-profile-66bb9c8b9d-86bpf        1/1     Running            0               8m
memcached-rate-68585796cb-gtpg6           1/1     Running            0               7m57s
memcached-reserve-646c7dcd6b-vwz8t        1/1     Running            0               7m52s
mongodb-geo-5ff578bcb8-bg6cr              1/1     Running            0               8m1s
mongodb-profile-758cb77f9f-thlpl          1/1     Running            0               8m
mongodb-rate-56cc8659c9-7dwkw             0/1     Pending            0               7m56s
mongodb-recommendation-55699465f7-f9hng   0/1     Pending            0               7m54s
mongodb-reservation-5994859869-zvvmt      1/1     Running            0               7m50s
mongodb-user-969c6c449-86tnh              0/1     Pending            0               7m48s
profile-676fcc86d5-v7c55                  0/1     Error              6 (3m16s ago)   7m59s
rate-c9bc58c85-klcq7                      0/1     CrashLoopBackOff   6 (25s ago)     7m55s
recommendation-8675c6c9ff-hqbl8           0/1     CrashLoopBackOff   6 (26s ago)     7m54s
reservation-7699b8fdd4-756lh              0/1     CrashLoopBackOff   5 (2m58s ago)   7m50s
search-65df96d8df-6lnw2                   1/1     Running            0               7m49s
user-8477d787d8-d8d45                     0/1     Error              6 (3m1s ago)    7m47s
\end{agenttool}

\begin{agentthought}{Mitigation}
The "geo" pod is in a CrashLoopBackOff state, and several other pods are pending or experiencing errors. This suggests there may be underlying issues with the application configuration or resource constraints.

I will check the logs of the "geo" pod to identify any specific errors causing the CrashLoopBackOff state.

\vspace{1em} \hrule \vspace{1em}

Call get\_logs with \{"namespace": "test-hotel-reservation", "service": "geo"\}
\end{agentthought}

\begin{agenttool}{get\_logs}{Mitigation}
{"level":"info","time":"2025-05-14T01:42:48Z","message":"TLS disabled."}
{"level":"info","time":"2025-05-14T01:42:48Z","message":"Set global log level: info"}
{"level":"info","time":"2025-05-14T01:42:48Z","message":"Tune: setGCPercent to 100"}
2025-05-14T01:42:48Z INF cmd/geo/main.go:23 > Reading config...
2025-05-14T01:42:48Z INF cmd/geo/main.go:36 > Read database URL: mongodb-geo:27017
2025-05-14T01:42:48Z INF cmd/geo/main.go:37 > Initializing DB connection...
2025-05-14T01:43:00Z PNC cmd/geo/db.go:26 > no reachable servers
panic: no reachable servers

goroutine 1 [running]:
github.com/rs/zerolog.(*Logger).Panic.func1({0x96e211, 0x0})
	/go/src/github.com/harlow/go-micro-services/vendor/github.com/rs/zerolog/log.go:405 +0x2d
github.com/rs/zerolog.(*Event).msg(0xc00019e150, {0x96e211, 0x14})
	/go/src/github.com/harlow/go-micro-services/vendor/github.com/rs/zerolog/event.go:158 +0x2b8
github.com/rs/zerolog.(*Event).Msg(...)
	/go/src/github.com/harlow/go-micro-services/vendor/github.com/rs/zerolog/event.go:110
main.initializeDatabase({0xc000028288, 0xc0000fddd0})
	/go/src/github.com/harlow/go-micro-services/cmd/geo/db.go:26 +0x167
main.main()
	/go/src/github.com/harlow/go-micro-services/cmd/geo/main.go:38 +0x877

Please take the next action
\end{agenttool}

\begin{agentanswer}{Mitigation}{Mitigation}
Root Cause:
1. The "frontend" service's "recommendation" operation is experiencing connectivity issues with the Consul service, indicated by the error message "connection refused." The Consul service was restarted to resolve any transient connectivity issues.
2. The "geo" operation in the "search" service is unable to bind the "geo-pvc" due to the persistent volume controller not recognizing the storage class "geo-storage." The missing storage classes were created to resolve this issue.
3. The "geo" pod is in a CrashLoopBackOff state due to a database connection error, specifically "no reachable servers." This indicates a problem with the MongoDB connection.

Mitigation Plan:
1. For the "frontend" service:
   - Ensure the frontend service is correctly configured to connect to the Consul service on port 8500.
   - Verify there are no network policies or firewall rules blocking traffic between the frontend and Consul services.
   - Monitor the Consul service to ensure it remains operational and accepting connections.

2. For the "geo" operation in the "search" service:
   - Ensure the persistent volume claims are bound correctly now that the storage classes have been created.
   - Investigate the MongoDB connection issue causing the "geo" pod to crash. Verify the MongoDB service is running and accessible at the expected endpoint.

3. For the pending PVCs:
   - Monitor the provisioner to ensure it completes the task of binding the remaining PVCs.
   - Check for any resource constraints or scheduling issues that may be preventing the PVCs from being bound.

Conclusion:
The primary issues have been addressed by restarting the Consul service and creating the missing storage classes. The remaining tasks involve monitoring the system to ensure the changes take effect and investigating the MongoDB connection issue for the "geo" pod.
\end{agentanswer}

\begin{tcolorbox}[frame hidden, title={Note (not part of the trajectory)}]
  \scriptsize 
  Although not all the pods are ready in the last observation, this doesn't means the agent failed this attempt.
  The operations it performed are correct, but the system is still in the process of stabilizing.
  Kubernetes will attempt to restart faulty pods periodically.
  After a few minutes of waiting, all pods will be ready and the workloads will be able to run without issues.
\end{tcolorbox}

\begin{frameworklog}{Validation \& Retry}{2}
Waiting 300.0 seconds for changes to take effect...
...
-----------------------------------------------------------------------
  117 requests in 10.00s, 48.95KB read
Requests/sec:     11.70  
Transfer/sec:      4.89KB
\end{frameworklog}

\begin{frameworklog}{AIOpsLab Validation}{2}
== Evaluation ==
Results:
{'success': True}
\end{frameworklog}

\let\clearpage\relax
\end{document}